\documentclass[12pt]{JHEP3}%
\usepackage{amssymb}
\usepackage{amsmath}
\usepackage{amsfonts}

\author{Lars Grant $^{a}$, Liat Maoz $^{b}$, Joseph Marsano $^{a}$,}
\author{Kyriakos Papadodimas $^{a}$ and Vyacheslav S.~Rychkov $^{b}$\\
\\ $^{a}$ Department of Physics, Harvard University \\
17 Oxford Street, Cambridge MA 02138, USA
\\ $^{b}$ Institute for Theoretical Physics, University of Amsterdam
\\ Valckenierstraat 65, 1018XE Amsterdam, The Netherlands\\
\\
E-mail: \email{lgrant@fas.harvard.edu},
\email{lmaoz@science.uva.nl}, \email{marsano@fas.harvard.edu},
\email{papadod@fas.harvard.edu},\\\email{rychkov@science.uva.nl}}

\title{Minisuperspace Quantization of ``Bubbling AdS" and Free Fermion Droplets}

\preprint{ITFA-2005-17}

 \abstract{We quantize the space of 1/2 BPS
configurations of Type IIB SUGRA found by Lin, Lunin and Maldacena
(hep-th/0409174), directly in supergravity. We use the
Crnkovi\'c-Witten-Zuckerman covariant quantization method to write
down the expression for the symplectic structure on this entire
space of solutions. We find the symplectic form explicitly around
$AdS_{5}\times S^{5}$ and obtain a $U(1)$ Kac-Moody algebra, in
precise agreement with the quantization of a system of $N$ free
fermions in a harmonic oscillator potential, as expected from
AdS/CFT. As a cross check, we also perform the quantization around
$AdS_{5}\times S^{5}$ by another method, using the known spectrum
of physical perturbations around this background and find precise
agreement with our previous calculation.}

\begin{document}

\section{Introduction}

Ever since the advent of AdS/CFT \cite{AdSCFT}, particular attention has
been paid to the 1/2 BPS sector which, on account of the high degree of
supersymmetry, is relatively simple and can often serve as a bridge
connecting the gauge theory and supergravity regimes. In fact, it has been
argued that this sector can be consistently decoupled \cite%
{Corley,Berenstein}, resulting in a system that admits a
description in terms of free fermions moving in a harmonic
oscillator potential. This has been well understood from the gauge
theory point of view, where the Lagrangian of the decoupled theory
is seen to be that of a complex matrix of oscillators whose
eigenvalues acquire Fermi-Dirac statistics from the integration
measure.

As a result of this, one expects a similar fermionic description of the 1/2
BPS sector from the gravity point of view. How such a description might
arise has been made clear in a recent paper by Lin, Lunin and Maldacena \cite%
{LLM}, who obtained all classical 1/2 BPS solutions and demonstrated that
they are naturally parametrized by planar droplets of various shapes. LLM
conjectured that these droplets should be identified with phase space
droplets describing semiclassical many-fermion states. In addition, \cite%
{LLM} showed that the quantization of flux matches with the quantization of
phase space area that one would expect from the fermion point of view.

We would like to take this analysis further. Is it possible to derive the
full quantum structure of the fermion system directly from the LLM
solutions? Such a result would be very much in line with the general spirit
of AdS/CFT.\footnote{%
This problem has also been recently studied by Mandal
\cite{Mandal}, who used the D3-brane probe method to derive an
action consistent with the fermionic description. In contrast, we
would like to understand the correspondence between the LLM
solutions and free fermion droplets directly from supergravity,
i.e.\ without the recourse to the microscopic description of the
LLM geometries in terms of D3 branes. This problem was also
discussed in \cite{Mandal}, and some suggestions concerning a
possible resolution were made.} The space of 1/2 BPS solutions is
simple enough that it seems plausible to address this question by
a direct quantization, at least in the limit of large $N$ where an
analysis within supergravity can be trusted. The usual problems
associated with canonical quantization in gravity are avoided in
this case because we are quantizing a \emph{space of solutions},
all of which automatically satisfy
the Hamiltonian constraint that is often so troubling{\footnote{%
Note also that, for this very reason, the quantization that we perform is
not a moduli space quantization as we do not consider fluctuations that take
us off the space of solutions in question.}}.

As a step in this direction, we first consider the sector of small 1/2 BPS
fluctuations about the $AdS_5\times S^5$ background and construct the
corresponding Hilbert space at large $N$. Performing the quantization in
this case is simple since the spectrum of fluctuations is known \cite{Kim}
and an effective action for this sector has previously been written down
\cite{Shiraz}. Such analysis, however, is not sufficiently general to deal
with fluctuations of more complicated backgrounds.

Fortunately, there exists a more general approach to quantization which does
not require explicit knowledge of the spectrum. This approach takes as its
starting point the symplectic form of Type IIB SUGRA, which encodes the
commutation relations that must be imposed on the system. The restriction of
this symplectic form to the LLM family of solutions can be computed
explicitly and defines a symplectic structure that enables us to perform our
quantization and study the Hilbert space about any background.

The symplectic form of supergravity can be most directly obtained by putting
this theory in the canonical form, analogous to the ADM construction for
pure gravity \cite{ADM}. This formalism is in essence non-covariant, since
it requires a particular space+time splitting of the metric, defining
canonical momenta etc. In practice, it is more convenient to use an
equivalent \textit{covariant} method of computing the symplectic form, which
was first proposed by Crnkovi\'{c} and Witten \cite{CW} and by Zuckerman
\cite{Zuckerman}. In this method one computes the symplectic form as an
integral of a symplectic current, which can be derived directly from the
action of the theory. In \cite{CW}, explicit expressions for the symplectic
currents for the Yang-Mills theory and for pure general relativity were
given. It is simple to generalize these results to Type IIB supegravity.
This permits us, in principle, to quantize fluctuations not only about $%
AdS_5\times S^5$, but also about more complicated backgrounds. After
demonstrating that this procedure gives results consistent with those of the
direct approach in the case of the $AdS_5\times S^5$ background, we then set
up a general formalism to do precisely this.

Finally, one might worry about the validity of this method of quantization,
which corresponds to a minisuperspace\footnote{%
This terminology \cite{Misner} has its roots in the concept of Wheeler's
\textit{superspace }\cite{Wheeler}, which is the space of all spatial
geometries in which geometrodynamics develops. Minisuperspace quantization
has been applied to models with finitely \cite{DeWitt, Misner} as well as
infinitely \cite{Kuchar} many degrees of freedom.} approximation of type IIB
supergravity where all degrees of freedom transverse to the space of 1/2 BPS
configurations are artificially frozen out. However, in the case at hand we
expect that our analysis indeed produces the correct Hilbert space and
spectrum. That we can neglect $\alpha^{\prime}$ corrections follows because
the spectrum is protected by supersymmetry and hence cannot depend on any
continuous parameter{\footnote{%
One can attempt to demonstrate this more formally by noting that the 1/2 BPS
spectrum can be computed as the limit of an index that is currently under
investigation \cite{KMM}.}}. That we can take the minisuperspace
approximation within supergravity is justified because all modes transverse
to the space of 1/2 BPS configurations decouple in the limit of large $N$.

The outline of this paper is as follows. In section 2, we review the 1/2 BPS
solutions of \cite{LLM} and their parametrization by planar droplets. We
also review the quantization of the relevant fermion system. In section 3,
we use the effective action of \cite{Shiraz} to quantize 1/2 BPS
fluctuations about $AdS_5\times S^5$. We then review the CWZ formalism in
section 4 and demonstrate its use by applying it to the same system of 1/2
BPS fluctuations about $AdS_5\times S^5$ in section 5. In section 6, we
initiate the program of quantizing fluctuations about backgrounds
corresponding to general droplets. We then make some concluding remarks in
section 7. Several calculational details have been deferred to the
Appendices.

\section{The LLM configurations}

\label{LLM}

\subsection{Supergravity solutions}

In \cite{LLM}, all regular 1/2 BPS solutions of Type IIB SUGRA with $%
SO(4)\times SO(4)\times \mathbb{R}$ symmetry were found. They have constant
dilaton and axion and vanishing 3-form. The explicit form of the solutions
is \footnote{%
We use the notation $F_{5}=\sum_{i_{1}<\ldots<i_{5}}F_{i_{1}\ldots
i_{5}}dx^{i_{1}}\wedge\ldots\wedge dx^{i_{5}}$, $|F_{5}|^{2}=\sum_{i_{1}<%
\ldots<i_{5}}F_{i_{1}\ldots i_{5}}F^{i_{1}\ldots i_{5}}.$}:%
\begin{align}
ds^{2} &
=-h^{-2}(dt+V_{i}dx^{i})^{2}+h^{2}(dy^{2}+dx^{i}dx^{i})+ye^{G}d%
\Omega_{3}^{2}+ye^{-G}d\tilde{\Omega}_{3}^{2}\,,  \label{lf} \\
F_{5} & =F\wedge d\Omega_{3}+\tilde{F}\wedge d\tilde{\Omega}_{3}\,,
\label{ls} \\
F & =dB,\qquad B=B_{t}(dt+V)+\hat{B}\,, \\
\tilde{F} & =d\tilde{B},\qquad\tilde{B}=\tilde{B}_{t}(dt+V)+\hat{\tilde{B}}%
\,,   \label{ll}
\end{align}
where $i=1,2$ , $d\Omega_{3}^{2}$ and $d\tilde{\Omega}_{3}^{2}$ are the
metrics on two unit 3-spheres $S^{3},\tilde{S}^{3}$, while $d\Omega_{3}$ and
$d\tilde{\Omega}_{3}$ are the corresponding volume forms. All the unknown
functions in (\ref{lf})-(\ref{ll}) depend only on $y,x_{1},x_{2}$ and are
fixed in terms of one function $Z(x_{1},x_{2}),$ which can only take the
values $\pm\frac{1}{2}.$ Namely, we have \footnote{%
We use here the standard notation for two-dimensional convolution: $f\ast
g(x)=\int f(x-x^{\prime })g(x^{\prime})d^{2}x^{\prime}$}:%
\begin{align}
z & =\frac{1}{\pi}\frac{y^{2}}{(x^{2}+y^{2})^{2}}\ast Z\,,  \label{z} \\
V_{i} & =\frac{\varepsilon_{ij}}{\pi}\frac{x_{j}}{(x^{2}+y^{2})^{2}}\ast Z\,,
\label{V} \\
h^{-2} & =\frac{y}{\sqrt{1/4-z^{2}}}\,,  \label{h} \\
e^{2G} & =\frac{1/2+z}{1/2-z}\,,   \label{G}
\end{align}%
\begin{align}
B_{t} & =-\frac{y^{2}}{4}e^{2G}, & \tilde{B}_{t} & =-\frac{y^{2}}{4}%
e^{-2G}\,, \\
d\hat{B} & =-\frac{y^{3}}{4}\ast_{3}d\left( \frac{z+1/2}{y^{2}}\right) , & d%
\hat{\tilde{B}} & =-\frac{y^{2}}{4}\ast_{3}d\left( \frac{z-1/2}{y^{2}}%
\right) ,   \label{bhat}
\end{align}
where $\ast_{3}$ is the flat space epsilon symbol in three dimensions $%
y,x_{1},x_{2}$.

The one-forms $B,\tilde{B}$ are defined up to a gauge transformation, and
can be found by solving the differential equations (\ref{bhat}) for $\hat{B}$%
, $\hat{\tilde{B}}$. One particular solution (arising in the gauge $\hat{B}%
_{y}=\hat{\tilde{B}}_{y}=0)$ is:
\begin{align}
& B_{i}=-\frac{y^{2}}{4({\frac{1}{2}}-z)}V_{i}-\frac{\varepsilon_{ij}}{4\pi }%
\frac{x_{j}}{x^{2}+y^{2}}\ast Z+\frac{1}{4}x_{1}\delta_{i,2}\,,  \notag \\
& \tilde{B}_{i}=-\frac{y^{2}}{4({\frac{1}{2}}+z)}V_{i}-\frac{\varepsilon
_{ij}}{4\pi}\frac{x_{j}}{x^{2}+y^{2}}\ast Z-\frac{1}{4}x_{1}\delta _{i,2}\,,
\label{gauge} \\
& B_{y}=\tilde{B}_{y}=0\,.  \notag
\end{align}

The function $Z(x_{1},x_{2})$ defines `droplets' on the $y=0$ plane. For a
droplet of finite size, the spacetime asymptotically approaches $%
AdS_{5}\times S^{5}$ with the (common) radius $R$ related to the area $A$ of
the droplet by \cite{LLM}
\begin{equation}
A=\pi R^{4}.
\end{equation}
On the other hand, the asymptotic radius is related to the number $N$ of
D3-branes making up the configuration by the standard relation (see e.g.
\cite{KlebanovRev})
\begin{equation}
R^{4}=\frac{\kappa_{10}N}{2\pi^{5/2}}.
\end{equation}
This shows that the total area of the droplet must be quantized\footnote{%
The relation $\kappa_{10}=8\pi^{7/2}\ell_{P}^{4}$ is useful in comparing
some of our equations to \cite{LLM}.}:%
\begin{equation}
\frac{A}{N}=\frac{\kappa_{10}}{2\pi^{3/2}}   \label{A/N}
\end{equation}
In the situation when several droplets are present, it was shown in \cite%
{LLM} using quantization of $F_{5}$-flux that the area of each droplet must
be quantized in the same units:
\begin{equation}
A_{i}=\frac{\kappa_{10}}{2\pi^{3/2}}N_{i},\qquad\sum N_{i}=N.
\end{equation}
In this paper we will only be considering one-droplet configurations.

\subsection{Dual description in terms of fermions}

As mentioned above, the LLM solutions corresponding to finite-size
droplets are asymptotically $AdS_{5}\times S^{5}$. One can then
make use of the AdS/CFT correspondence and relate these gravity
solutions to $\mathcal{N}=4$ super Yang-Mills on $S^{3}\times
\mathbb{R}$. As the geometries are all 1/2 BPS, the relevant
operators on the Yang-Mills side are chiral primary operators with
conformal weight equal to their $U(1)$ R-charge: $\Delta=J$. It
was argued \cite{Corley,Berenstein} that this sector of
$\mathcal{N}=4$ super Yang-Mills is actually a $U(N)$ one-matrix
quantum mechanical system with a harmonic oscillator potential.

There are two equivalent ways of looking at this model. One, the `closed
string' picture, is a description of the system in terms of $N^{2}$ free
harmonic oscillators with the Hamiltonian
\begin{equation}
H = \frac{1}{2}(a^{\dagger})^{i}_{j}a^{j}_{i}+\frac{1}{2}N^{2}\,,
\end{equation}
where $(a^{\dagger})^{i}_{j} , a^{i}_{j}$ are $N^{2}$ creation and
annihilation operators, obeying the commutators:
\begin{equation}
[(a^{\dagger})^{i}_{j},a^{k}_{l}]=\delta^{i}_{l}\delta^{k}_{j}\,.
\end{equation}
The states
\begin{equation}
Tr[(a^{\dagger})^{n_{1}}]\ldots Tr[(a^{\dagger})^{n_{k}}]|0\rangle
\label{closed}
\end{equation}
where $k\ge1$, and $n_{1},n_2,\ldots n_{k}$ is a set of non-increasing
integers between 1 and $N$, are a basis for the singlet states of the
system. Every such state can also be represented in terms of a $U(N)$ Young
tableaux.

Another way of looking at this model is through the `open string' picture.
In this picture the Hamiltonian is
\begin{equation}
H=\frac{1}{2}\sum_{i=1}^{N}-\partial_{\lambda_{i}}^{2}+\lambda_{i}^{2}\,,
\end{equation}
and it describes $N$ free fermions in the harmonic oscillator potential well
\cite{Berenstein,Sunny}. This can also be rewritten as a system of $N$
fermions in a constant magnetic field confined to the Lowest Landau Level
\cite{Berenstein,Ber2}. A basis of $N$-particle wave functions is given by
the Slater determinant of $N$ single particle wave functions:
\begin{equation}
\psi_{(n_{1},\ldots,n_{N})}(\lambda_{1},\ldots,\lambda_{N})\sim\sum_{\sigma%
\in S_{N}}sgn(\sigma)\prod_{i}\psi_{n_{i}}(\lambda_{\sigma(n_{i})})\,.
\end{equation}
The ground state corresponds to the fermions filling up the lowest $N$
levels.
We can describe excited states by a set of $N$ non-increasing integers $%
n_1,n_2,\ldots n_N$ including zero. We move the fermion in level $N$ to $%
N+n_1$, the one in $N-1$ to $N-1+n_2$ etc. To this state we can associate a
Young tableaux with rows of length $n_1,n_2,\ldots n_N$. The energy of this
state is $\sum_i n_i$. Any two states described by two different sequences
of integers $n_i$ are orthogonal. This completes the description of the
Hilbert space/spectrum of this system. It is easy to write the exact
partition function:
\begin{equation}
Z_{exact}(\beta) = \prod_{n=1}^N \frac{1}{1-e^{-\beta n}}\,.  \label{zexact}
\end{equation}
Writing this partition function in the form
\begin{equation}
Z_{exact}(\beta) = \sum_{E_i} g(E_i) e^{-\beta E_i}\,,
\end{equation}
we can read off the degeneracy $g(E_i)$ of the energy state $E_i$.

\subsection{Fermion quantization in the large $N$ limit}

\label{Fermions} We would like to setup a formalism to compare the
quantization of the LLM solutions with the corresponding super Yang-Mills
states. The correspondence is supposed to work in the large $N$ limit. On
the supergravity side we will compute the symplectic form using the CWZ
method, promote Poisson brackets to commutators and get a Hilbert space
description. On the SYM side we will base our treatment on the `open string'
description explained above. Thus we have a system of $N$ fermions in the
harmonic potential, for which we already have an exact Hilbert space
description. How does one pass to the large $N$ limit? In the large $N$
limit the states of the many-fermion system are well described as droplets
in the one-particle phase space. It is simpler to go first one step
backwards, and discuss the \textit{classical dynamics} of these droplets%
\footnote{%
This so-called hydrodynamic approach is standard in describing
edge states in the Quantum Hall Effect, see \cite{Wen} and
references therein. For a different approach to the large $N$
limit, using a noncommutative theory for Wigner phase space
density, see \cite{Dhar} and references therein.}. This discussion
will lead us to the corresponding symplectic form, and thus we
will be able to recover the large $N$ Hilbert space quantizing
this symplectic form. The final result, Eq.~(\ref{fermPB}), is
well known; in a more general setting it is derived in
\cite{Poly}.

The harmonic potential one-fermion Hamiltonian is
\begin{equation}
H=\frac{p^{2}+q^{2}}{2}.   \label{harm}
\end{equation}
We will describe the boundary of a droplet in polar coordinates:%
\begin{align}
p =r(\phi)\sin\phi\,,\quad q =r(\phi)\cos\phi\,,
\end{align}
where we assume that $r(\phi)$ is a single-valued function. The boundary of
the droplet evolves according to the one-fermion equation of motion, which
in this case is simply the clockwise rotation with unit angular velocity:%
\begin{equation}
\dot{p}=-q,\quad\dot{q}=p.
\end{equation}
This implies that the boundary $r(\phi)$ evolves in time according to the
simple equation%
\begin{equation}
\dot{r}=r^{\prime}.   \label{rot}
\end{equation}
We would now like to find a symplectic form on the phase space of droplets,
so that the Hamilton equation%
\begin{equation}
\dot{r}=\{r,H_{tot}\}
\end{equation}
would coincide with (\ref{rot}). Here $H_{tot}$ is the total energy of the
droplet state, given by the integral of the one-particle Hamiltonian:%
\begin{equation}
H_{tot}=\int\int_{droplet}\frac{dp\,dq}{2\pi\hbar}\frac{p^{2}+q^{2}}{2}\,.
\end{equation}
(This formula takes into account that one state occupies a phase
space area of $2\pi\hbar$ in the semiclassical description.)
Notice that this $\hbar$ is the effective 1d Planck constant
(which has nothing to do with the 10d Planck constant
$\hbar_{10}$. In fact we put $\hbar_{10}=1$, as usual). This
constant can be determined from%
\begin{equation}
A=2\pi\hbar N \,,   \label{tofindh}
\end{equation}
where $A$ is the area of the droplet and $N$ is the total number of fermions
(which is the same as the number of D3 branes in the gravity description).
The total energy is easily computed in terms of $r(\phi)$:%
\begin{equation}
H_{tot}=\frac{1}{16\pi\hbar}\int d\phi\,r^{4}(\phi)\equiv\frac{1}{16\pi\hbar
}\int d\phi\,f^{2}(\phi),   \label{Sug}
\end{equation}
where we defined $f(\phi)\equiv r^{2}(\phi).$ Let us define the following
Poisson bracket (see \cite{Poly})%
\begin{equation}
\{f(\phi),f(\tilde{\phi})\}=8\pi\hbar\,\delta^{\prime}(\phi-\tilde{\phi}).
\label{fermPB}
\end{equation}
The Hamilton equation corresponding to this bracket is
\begin{equation}
\dot{f}=\{f,H_{tot}\}=f^{\prime}.
\end{equation}
This equation is equivalent to (\ref{rot}), and thus we conclude that (\ref%
{fermPB}) is the correct Poisson bracket.

Although we derived this Poisson bracket for the particular one-fermion
Hamiltonian (\ref{harm}), it is in fact completely general and will describe
the motion of droplets of noninteracting fermions described by an arbitrary
one-particle Hamiltonian $H(p,q)$. The total energy in such a general case is%
\begin{equation}
H_{tot}=\int\frac{dpdq}{2\pi\hbar}H(p,q)=\frac{1}{2\pi\hbar}\int d\phi\int
_{0}^{r(\phi)}dr\,rH(r\sin\phi,r\cos\phi)
\end{equation}
The evolution of the boundary of the droplet computed using the Poisson
bracket (\ref{fermPB}) is
\begin{align}
\dot{f} & =8\pi\hbar\int d\tilde{\phi}\,\delta^{\prime}(\phi-\tilde{\phi })%
\frac{\delta H_{tot}}{\delta f(\tilde{\phi})}=8\pi\hbar\frac{d}{d\phi }%
\left( \frac{\delta H_{tot}}{\delta f(\phi)}\right)  \notag \\
& =2\frac{d}{d\phi}\,[H(r(\phi)\sin\phi,r(\phi)\cos\phi)]  \notag \\
& =2\left[ (r^{\prime}\sin\phi+r\cos\phi)H_{p}+(r^{\prime}\cos\phi-r\sin
\phi)H_{q}\right] \,.   \label{toagree}
\end{align}
We have to compare this with the evolution of the boundary which is
generated by the one-fermion Hamilton equations%
\begin{equation}
\dot{q}=H_{p}\qquad\dot{p}=-H_{q}.
\end{equation}
After a small time increment $dt$ we have%
\begin{align}
dr^{2} & =2pdp+2qdq=\left[ 2qH_{p}-2pH_{q}\right] dt\,, \\
d\phi & =r^{-2}[qdp-pdq]=-r^{-2}\left[ qH_{q}+pH_{p}\right] dt\,, \\
dr^{2}|_{\phi=const} & =dr^{2}-(r^{2})^{\prime}d\phi  \notag \\
& =\left[ 2qH_{p}-2pH_{q}+2(r^{\prime}/r)(qH_{q}+pH_{p})\right] dt  \notag \\
& =2[(q+r^{\prime}\sin\phi)H_{p}+(r^{\prime}\cos\phi-p)H_{q}]dt,
\end{align}
which agrees with (\ref{toagree}) as it should$\footnote{%
It should be noted that for a general Hamiltonian a region described at the
initial moment of time $t=0$ by a single-valued function $r(\phi)$ can
evolve at a later moment of time into a region for which $r(\phi)$ is
multiple-valued. Still near $t=0$ we can describe dynamics in terms of the
Poisson bracket (\ref{fermPB}).}$.

Now that we determined the Poisson bracket (\ref{fermPB}) generating
classical dynamics, we can immediately write down the corresponding
commutation relation multiplying by $i\hbar.$ These commutation relations
will have to be reproduced from the gravity side. Thus the result we should
aim for is
\begin{equation}
\lbrack f(\phi),f(\tilde{\phi})]=i8\pi\hbar^{2}\delta^{\prime}(\phi -\tilde{%
\phi})\equiv i\frac{\kappa_{10}^{2}}{2\pi^{4}}\delta^{\prime}(\phi-\tilde{%
\phi}).   \label{ffcomm}
\end{equation}
where we expressed $\hbar$ via $\kappa_{10}$ using (\ref{tofindh}), (\ref%
{A/N}).

We finish this section by showing how to construct a Hilbert space
realization of (\ref{ffcomm}). We start by expanding $f(\phi)$ in Fourier
series:%
\begin{equation}
f(\phi)=\sum f_{n}e^{in\phi},\qquad f_{-n}=f_{n}^{\ast}.   \label{fexp}
\end{equation}
The zero mode is fixed in terms of the droplet area:%
\begin{equation}
f_{0}=\frac{A}{\pi}.
\end{equation}
The $n\neq0$ modes correspond to the area-preserving deformations.
Substituting (\ref{fexp}) in (\ref{ffcomm}), we get commutation relation
between the corresponding Fourier coefficients:%
\begin{equation}
\lbrack f_{n},f_{m}]=-\frac{\kappa_{10}^{2}}{4\pi^{5}}n\delta_{n+m}
\label{torepr}
\end{equation}
This is just a $U(1)$ Kac-Moody algebra\footnote{%
One can also add to the $f_{n}$'s an energy-momentum tensor using the
Sugawara construction. It turns out to be proportional to (\ref{Sug}). Thus $%
f(\phi)$ is related to a free chiral boson $X(\phi)$ through $%
f(\phi)\sim\partial_{\phi}X(\phi)$, as can also be seen from (\ref{ffcomm}).
In the context of the Quantum Hall Effect, this anomaly equation expresses
the non-conservation of the charge density on the boundary in presence of an
external electric field \cite{Haldane}.}. The Hilbert space of the theory
can be constructed as a bosonic Fock space with annihilation and creation
operators $c_{n},c_{n}^{\dagger}$ where%
\begin{equation}
c_{n}=\left( \frac{\kappa_{10}^{2}}{4\pi^{5}}n\right) ^{-1/2}f_{-n},\qquad
n>0.
\end{equation}
Of course there is no contradiction with the fermionic statistics of the
individual particles: the droplet exitations are collective modes and are
free to satisfy any statistics.

Finally, as a check of the validity of this semiclassical description we can
compute the partition function of many fermions in a harmonic potential
using the semiclassical method and compare with the result of the exact
quantization. Using the bosonic Hilbert space constructed above by the
operators $c_n,c_n^\dagger$ we find:
\begin{equation}
Z_{semiclassical}(\beta)= \prod_{n=1}^\infty \frac{1} {1-e^{-\beta n}}\,.
\end{equation}
Comparing with (\ref{zexact}) we see that the semiclassical method
reproduces the correct spectrum of the theory $E_i=1,2,3,\ldots$ and the
exact degeneracies of these states provided that $E<N$. For higher energies
the degeneracies predicted by the semiclassical quantization do not agree
with those of the exact quantization. But since our calculations in gravity
are valid only for large $N$, we will always be in the regime where the
semiclassical approximation is reliable.

\section{The $AdS$ background I -- Effective action approach}

\label{AdSOne} In this section, we consider small 1/2 BPS fluctuations about
the $AdS_{5}\times S^{5}$ background and quantize them directly via an
effective action obtained from Type IIB SUGRA. This is not too difficult to
do as the spectrum of modes on $AdS_{5}\times S^{5}$ is well known \cite{Kim}
and, moreover, the effective action for the 1/2 BPS fluctuations has been
worked out to the order in which we are interested by \cite{Shiraz}. Thus,
to compare with the fermion quantization of the previous section, we need
only to relate the LLM parametrization of 1/2 BPS fluctuations in terms of
deformations of a droplet to the modes of \cite{Kim}, which differ by a
diffeomorphism. We review the LLM description of $AdS_{5}\times S^{5}$ and
1/2 BPS fluctuations about this background and connect to the modes of \cite%
{Kim} in the first subsection. We then utilize the effective action of \cite%
{Shiraz} to perform the quantization. The result of this quantization, Eq.~(%
\ref{acrs}),\ perfectly agrees with what we expect from the fermion side,
Eq.~(\ref{torepr}).

\subsection{LLM modes about $AdS_{5}\times S^{5}$}

\label{modematching}

In the language of LLM, the $AdS_{5}\times S^{5}$ background corresponds to
a circular droplet in the $(x_{1},x_{2})$ plane with radius $%
r_{0}=R_{AdS}^{2}$ that we set to 1 by convention. Evaluating the various
functions appearing in the metric for such a droplet and using polar
coordinates $r,\phi$ on the $(x_{1},x_{2})$ plane, we find
\begin{equation}
\begin{split}
z(r,y) & =\frac{r^{2}-1+y^{2}}{2\sqrt{(r^{2}+1+y^{2})-4r^{2}}}\,, \\
V_{r} & =0\,, \\
V_{\phi} & =\frac{1}{2}\left( \frac{r^{2}+1+y^{2}}{\sqrt{%
(r^{2}+1+y^{2})^{2}-4r^{2}}}-1\right) \,.
\end{split}%
\end{equation}
If we make the change of coordinates%
\begin{equation}
\begin{split}
y & =\sinh\rho\sin\theta\,, \\
r & =\cosh\rho\cos\theta\,, \\
\phi & =\tilde{\phi}+t\,.
\end{split}
\label{globalads}
\end{equation}
then the LLM metric (\ref{lf}) becomes that of $AdS_{5}\times S^{5}$ in
global coordinates%
\begin{equation}
ds^{2}=-\cosh^{2}\rho\,dt^{2}+d\rho^{2}+\sinh^{2}\rho\,d\Omega_{3}^{2}+d%
\theta^{2}+\cos^{2}\theta\,d\tilde{\phi}^{2}+\sin^{2}\theta\,d\tilde{\Omega }%
_{3}^{2}\,,   \label{glob}
\end{equation}
and the five-form becomes%
\begin{equation}
F_{5}=\cosh\rho\sinh^{3}\rho\,dt\wedge d\rho\wedge d\Omega_{3}+\cos\theta
\sin^{3}\theta\,d\theta\wedge d\tilde{\phi}\wedge d\tilde{\Omega}_{3}\,.
\end{equation}

We now consider perturbations of this background corresponding to small
ripples of the droplet. The boundary of the perturbed droplet in polar
coordinates is given by $r(\phi)=1+\delta r(\phi)$. We expand $\delta r(\phi)
$ in Fourier series\footnote{%
The zero mode is absent, since the deformation should be area preserving (to
the first order in $\delta r$).}:%
\begin{equation}
\delta r(\phi)=\sum_{n\neq0}a_{n}e^{in\phi},\qquad a_{n}^{\ast}=a_{-n}.
\label{droppar}
\end{equation}
Using (\ref{z}), (\ref{V}), it is easy to find the first order shifts in the
functions $z,V_{i}$ due to the presence of these ripples:%
\begin{equation}
\begin{split}
\delta z(r,\phi,y) & =-\sum_{n}\frac{2y^{2}a_{n}e^{in\phi}}{B^{2}}\left(
\frac{1+|n|\sqrt{1-A^{2}}}{(1-A^{2})^{3/2}}\right) \left[ \frac {1-\sqrt{%
1-A^{2}}}{A}\right] ^{|n|}\,, \\
\delta V_{r}(r,\phi,y) & =\sum_{n}\frac{2ina_{n}e^{in\phi}}{B^{2}A\sqrt{%
1-A^{2}}}\left[ \frac{1-\sqrt{1-A^{2}}}{A}\right] ^{|n|}\,, \\
\delta V_{\phi}(r,\phi,y) & =\sum_{n}\frac{2ra_{n}e^{in\phi}}{%
B^{2}(1-A^{2})^{3/2}}\left[ (r-A)+(r-A^{-1})|n|\sqrt{1-A^{2}}\right] \left[
\frac{1-\sqrt{1-A^{2}}}{A}\right] ^{|n|}\,,
\end{split}
\label{dllmfcns}
\end{equation}
where%
\begin{equation}
\begin{split}
A & \equiv\frac{2r}{r^{2}+y^{2}+1}\,, \\
B & \equiv r^{2}+y^{2}+1\,.
\end{split}
\label{ABdef}
\end{equation}
Inserting these variations into (\ref{lf}), we obtain the metric
perturbations which correspond to the excitation of ripples on the droplet
in the fermi picture. To identify these perturbations with the modes of Kim
\textit{et al.}\cite{Kim}, two additional operations are required. First of
all, we have to move to global $AdS$ coordinates via (\ref{globalads}). In
this way we arrive at a certain metric perturbation $h_{mn}$, which is still
not in the Kim \textit{et al.} form. To achieve total coincidence, we have
to perform an additional linear gauge transformation%
\begin{equation}
h_{mn}\rightarrow h_{mn}-\left( \nabla_{m}\xi_{n}+\nabla_{n}\xi_{m}\right)
\label{lingtr}
\end{equation}
for the particular choice of $\xi_{i}$ with nonzero components%
\begin{equation}
\begin{split}
\xi_{t} & =-\sum_{n}\frac{in}{|n|}a_ne^{in(\tilde{\phi}+t)}\left( \frac{%
\cos\theta}{\cosh\rho}\right) ^{|n|}\,, \\
\xi_{\rho} & =-\sum_{n}\frac{2\cos^{2}\theta\tanh\rho}{\cosh(2\rho)-\cos(2%
\theta)}a_{n}e^{in(\tilde{\phi}+t)}\left( \frac{\cos\theta}{\cosh\rho }%
\right) ^{|n|}\,, \\
\xi_{\theta} & =\sum_{n}\frac{\sin(2\theta)}{\cosh(2\rho)-\cos(2\theta)}%
a_{n}e^{in(\tilde{\phi}+t)}\left( \frac{\cos\theta}{\cosh\rho}%
\right)^{|n|}\,.
\end{split}%
\end{equation}
The resulting metric perturbations then take the form%
\begin{equation}
h_{\mu\nu}=\sum_{n\neq0}\left( -\frac{6}{5}|n|s_{n}Y_{n}g_{\mu\nu}+\frac {4}{%
|n|+1}Y_{n}\nabla_{(\mu}\nabla_{\nu)}s_{n}\right) \,,\qquad
h_{\alpha\beta}=\sum_{n\neq0}2|n|s_{n}Y_{n}g_{\alpha\beta}\,,
\label{metpert}
\end{equation}
where $\mu,\nu$ run over the $AdS_{5}$ directions, $\alpha,\beta$ run over
the $S^{5}$ directions, $g_{mn}$ is the unperturbed $AdS_{5}\times S^{5}$
metric, the covariant derivatives are computed with respect to the metric $%
g_{mn}$, and%
\begin{align}
s_{n} & =\frac{|n|+1}{2|n|\cosh^{|n|}\rho}a_{n}e^{int}\,,  \notag \\
Y_{n} & =e^{in\tilde{\phi}}\cos^{|n|}\theta\,.   \label{snyn}
\end{align}
Written in this form, the metric perturbations can be identified with a
subclass of modes studied in \cite{Kim}, \cite{Shiraz}. Thus, the modes of
\cite{Kim}, \cite{Shiraz} have the same functional dependence as (\ref%
{metpert}) with functions $s_{n},$ $Y_{n}$ satisfying the differential
equations%
\begin{equation}
\nabla_{S^{5}}^{2}Y_{n}=-n(n+4)Y_{n}\,,\qquad%
\nabla_{AdS_{5}}^{2}s_{n}=n(n-4)s_{n}\,,   \label{difeq}
\end{equation}
so that, for instance, the $Y_{n}$ are $S^{5}$ spherical harmonics. Our $%
s_{n}$ and $Y_{n}$ correspond, not surprisingly, to the subclass of
solutions of these equations that are constant on $S^{3}$ and $\tilde{S}^{3}$%
.

The perturbation of the 4-form potential could be in principle analyzed
using the same method (expressing perturbations in LLM coordinates, passing
to the global coordinates, and following up by a gauge transformation).
However, this is unnecessary, because the end result of this computation can
be uniquely reconstructed from the knowledge of metric perturbations (\ref%
{metpert}). Namely, the 4-form potential perturbations can be written in the
gauge $\nabla^{\alpha}a_{\alpha mnp}=0$ as \cite{Kim}
\begin{align}
\delta a_{\alpha\beta\gamma\delta} & =-\epsilon_{\alpha^{\prime}\alpha
\beta\gamma\delta}s_{n}\nabla^{\alpha^{\prime}}Y_{n}\,,  \notag \\
\delta a_{\mu\nu\rho\lambda} & =\epsilon_{\mu^{\prime}\mu\nu\rho\lambda
}Y_{n}\nabla^{\mu^{\prime}}s_{n}\,,   \label{deltaa}
\end{align}
where $\epsilon_{\alpha^{\prime}\alpha\beta\gamma\delta}$ ($\epsilon
_{\mu^{\prime}\mu\nu\rho\lambda}$) is the curved-space $\epsilon$ symbol on $%
S^{5}$ ($AdS_{5}$).

\subsection{Effective action}

\label{Effective}

The effective second-order action describing perturbations of $AdS_{5}\times
S^{5}$ of the form (\ref{metpert}),(\ref{deltaa}) has been derived in \cite%
{Shiraz} and has the form%
\begin{equation}
S=\frac{1}{2\kappa_{10}^{2}}\sum_{I}\int\,d^{5}x\,\sqrt{-g_{AdS}}\frac{A_{I}%
}{2}\left[ -(\nabla s_{I})^{2}-n(n-4)(s_{I})^{2}\right] \,,   \label{effact}
\end{equation}
where the index $I$ labels spherical harmonics. This effective action was
only derived in \cite{Shiraz} for modes with $n\geq2$,\ where $n=n(I)$ is
the parameter in the eigenvalue equations (\ref{difeq}). As a result, we
leave the $n=1$ mode out of consideration in this section. It will however
be covered by the alternative treatment in Sec. \ref{AdSTwo}.

The action (\ref{effact}) was derived under the assumption that the
spherical harmonics are real and satisfy an orthogonality relation%
\begin{equation}
\int_{S^{5}}Y_{I}Y_{I^{\prime}}=z_{I}\delta_{I,I^{\prime}}\,.
\end{equation}
The action is obtained by expanding the Type IIB SUGRA action to the second
order in perturbations and performing the $S^{5}$ integration{\footnote{%
More precisely, the expansion is done in the 'actual' IIB SUGRA action \cite%
{actual} in which the selfduality constraint on the 5-form is enforced via
an auxiliary field.}}. The constants $A_{I}$ are given by:
\begin{equation}
A_{I}=32\frac{n(n-1)(n+2)}{n+1}z_{I}\,.
\end{equation}

In this subsection, we will apply this effective action to quantize the LLM
family of 1/2 BPS fluctuations. A minor complication arises since angular
momentum conservation prevents a direct restriction to the sector in
question at the level of the action so we will have to include modes that
rotate in the opposite direction. After quantization, though, it will be
easy to make the appropriate truncation of the quantized Hilbert space.

We now proceed using the real modes

\begin{align}
s_{n,1} & =\frac{n+1}{2n\cosh^{n}\rho}x_{n}(t)\,,\qquad Y_{n,1}=\cos
^{n}\theta\cos(n\phi)\,,  \notag \\
s_{n,2} & =\frac{(n+1)}{2n\cosh^{n}\rho}y_{n}(t)\,,\qquad Y_{n,2}=\cos
^{n}\theta\sin(n\phi)\,.   \label{real}
\end{align}
These modes are chosen because\ they are related to (\ref{snyn}):%
\begin{equation}
s_{n}Y_{n}+s_{-n}Y_{-n}=\sum_{i=1}^{2}s_{n,i}Y_{n,i}   \label{snynrew}
\end{equation}
for the particular choice of real functions $x_{n}(t)$ and $y_{n}(t)$%
\begin{equation}
x_{n}(t)=a_{n}e^{int}+c.c.,\qquad y_{n}(t)=ia_{n}e^{int}+c.c.   \label{xydef}
\end{equation}
Substituting (\ref{real}) into the effective action (\ref{effact}) and
performing the spatial integrations, we obtain the following effective
action for $x_{n}(t),y_{n}(t):$%
\begin{equation}
S=\frac{2\pi^{5}}{\kappa_{10}^{2}}\sum_{n\geq2}\int\,dt\,\left[ \frac {1}{%
n^{2}}\left( \dot{x}_{n}^{2}+\dot{y}_{n}^{2}\right) -\left(
x_{n}^{2}+y_{n}^{2}\right) \right] \,.   \label{xyact}
\end{equation}
The quantization of this system is now straightforward. We have the
following equal-time canonical commutators:%
\begin{equation}
\lbrack x_{n},\dot{x}_{m}]=[y_{n},\dot{y}_{m}]=in^{2}\left( \frac{\kappa
_{10}^{2}}{4\pi^{5}}\right) \delta_{n,m}\,.   \label{xycr}
\end{equation}
Representing $x_{n}(t),$ $y_{n}(t)$ as
\begin{equation}
x_{n}=c_{n}e^{int}+h.c.,\qquad y_{n}=d_{n}e^{int}+h.c.,
\end{equation}
we get the following nonzero commutation relations for $c_{n},d_{n}$%
\begin{equation}
\lbrack c_{n}^{\dagger},c_{m}]=[d_{n}^{\dagger},d_{m}]=\frac{\kappa_{10}^{2}%
}{8\pi^{5}}n\delta_{n,m}\,.   \label{cd}
\end{equation}
As discussed before, the full Hilbert space of the system \eqref{xyact}
contains both the 1/2-BPS fluctuations of interest as well as those carrying
opposite angular momentum. We thus have to truncate this Hilbert space
according to (\ref{xydef}). In order to do this consistently, we put
\begin{equation}
c_{n}=a_{n}+b_{n},\quad d_{n}=i(a_{n}-b_{n})\,.
\end{equation}
Then from (\ref{cd}) we get the following commutators for $a_{n},b_{n}:$%
\begin{equation}
\lbrack a_{n}^{\dagger},a_{m}]=[b_{n}^{\dagger},b_{m}]=\frac{\kappa_{10}^{2}%
}{16\pi^{5}}n\delta_{n,m}\,.   \label{ab}
\end{equation}
Since all commutators between $a_{n}$ and $b_{m}$ vanish, the $b_{n}$ sector
decouples and it is consistent to truncate the full Hilbert space keeping
only the states in the $a_{n}$ sector. In the classical theory this
corresponds to setting $b_{n}=0,$ which is consistent with (\ref{xydef}).
The commutation relation for $a_{n}$ in (\ref{ab}), which can be rewritten
as
\begin{equation}
\lbrack a_{n},a_{m}]=-\frac{\kappa_{10}^{2}}{16\pi^{5}}n\delta_{n+m}\,,
\label{acrs}
\end{equation}
is easily seen to be consistent with (\ref{torepr}) using the fact that $%
\delta(r^{2})=2\delta r,$ and thus $f_{n}=2a_{n}.$

Finally, we note that the Hamiltonian density for the system (\ref{xyact}), (%
\ref{xydef}) is also easy to compute:%
\begin{equation}
H=\frac{16\pi^{5}}{\kappa_{10}^{2}}\sum_{n\geq2}a_{n}a_{-n}\,,
\label{hamdens}
\end{equation}
and agrees with what one obtains from LLM (2.32).

\section{The CWZ method of minisuperspace quantization}

\label{CWmethod}

The quantization method described in the previous section does not seem to
be applicable for the general droplet case. As motivated in the
introduction, in this section we will review and apply to Type IIB SUGRA the
covariant phase space approach of Crnkovi\'{c}-Witten-Zuckerman. It is this
method that will be used in the rest of the paper.

\subsection{Generalities}

\label{Gen}Crnkovi\'{c} and Witten \cite{CW} and Zuckerman \cite{Zuckerman}
(see also \cite{Witten}) proposed an alternative approach to quantizing
Lagrangian theories which is equivalent to the usual Hamiltonian formalism
and is particularly suited for minisuperspace quantization. The starting
point of the CWZ approach is that, in a general field theory, the canonical
phase space is in one-to-one correspondence, and thus can be identified,
with the space of solutions of the classical field equations. Thus we should
be able to define the symplectic form needed for quantization directly on
the space of solutions, without recourse to some specific canonical
variables.

In this approach, the symplectic form is computed as an integral of a
specific \textit{symplectic current }$J^{l}$ over an initial hypersurface $%
\Sigma$:%
\begin{equation}
\omega=\int d\Sigma_{l}J^{l}.   \label{scalar}
\end{equation}
The simplest example is that of the free scalar field $\phi$ in Minkowski
space, in which the symplectic current is given by
\begin{equation}
J^{l}(x)=\partial^{l}\delta\phi(x)\wedge\delta\phi(x).   \label{scalarJ}
\end{equation}
Here, using the notation of \cite{CW}, $\delta\phi$ is an element of the
tangent space to the space of solutions. In particular, $\delta\phi$
satisfies the linearized equations of motion (which in this simplest linear
situation coincide with the original ones). By evaluating $\delta\phi$ at a
point $x,$ we get a one-form on the phase space, $\delta\phi(x).$
Analogously $\partial^{l}\delta\phi(x)$ is also a one-form for each $x.$
More forms can be constructed using the exterior product $\wedge$ and the
exterior differentiation $\delta.$ We see that $\omega$ defined by (\ref%
{scalar}) is a two-form on the phase space. Moreover, it is closed because $%
\delta(\delta \phi(x))=0.$ It is also invariant under variations of $\Sigma,$
because the $J^{l}$ as defined by (\ref{scalarJ}) is conserved: $%
\partial_{l}J^{l}=0.$ Note that one needs the equations of motion to show
this and hence it is true only when the symplectic form is evaluated on the
phase space of solutions.

To complete the free scalar quantization, one should choose a particular %
\hyphenation{pa-ra-met-ri-za-tion} parametrization of the solutions to the
Klein-Gordon equation, e.g.\ by expanding solutions into plane waves, and
compute the symplectic form (\ref{scalar}) in terms of the coefficients of
such an expansion. The Poisson brackets following from the computed
symplectic form are then promoted to commutators using the Dirac
prescription. The resulting quantization is completely equivalent to the
usual Hamiltonian approach.

It turns out that the above method is not limited to free scalar fields, but
can be adapted to a wide range of theories, including those with gauge
symmetries, where the symplectic form has to satisfy an additional
constraint of gauge invariance. In \cite{CW}, appropriate symplectic
currents were written down for Abelian and non-Abelian gauge theories as
well as for pure gravity; for recent applications of the method to other
physical theories see \cite{WZW}. For theories with several fields it
becomes easier to compute the symplectic current using the following general
method \cite{Zuckerman}, \cite{Wald}. We will assume that the Lagrangian $%
L=L(\phi_{A},\partial_{l}\phi_{A})$ (where the index $A$ numbers the fields)
does not contain second- and higher-order derivatives, so that the classical
equations of motion are
\begin{equation}
\frac{\partial L}{\partial\phi_{A}}-\partial_{l}\frac{\partial L}{%
\partial\partial_{l}\phi_{A}}=0.
\end{equation}
Under these conditions, the symplectic current is defined by{\footnote{%
Actually, what we define is a vector density and hence is appropriate to be
integrated without the usual $\sqrt{-g}$ factor.}}

\begin{equation}
J^{l}=\delta\left[ \frac{\partial L}{\partial\partial_{l}\phi_{A}}\right]
\delta\phi_{A} = \delta\left[ \frac{\partial L}{\partial\partial_{l}\phi_{A}}%
\delta\phi_{A}\right] .   \label{_current}
\end{equation}
(From now on we, following \cite{CW}, will usually omit the wedge product
sign, treating $\delta\phi$'s as anticommuting objects. Of course, wedge
products can be reinstated at any moment, if desired.) It is obvious from
the second representation that $J^{l}$ is closed: $\delta J^{l}=0$. It is
also conserved, due to the equations of motion:
\begin{align}
\partial_{l}J^{l} & =\delta\left[ \partial_{l}\frac{\partial L}{%
\partial\partial_{l}\phi_{A}}\delta\phi_{A}+\frac{\partial L}{\partial
\partial_{l}\phi_{A}}\delta\partial_{l}\phi_{A}\right]  \notag \\
& =\delta\left[ \frac{\partial L}{\partial\phi_{A}}\delta\phi_{A}+\frac{%
\partial L}{\partial\partial_{l}\phi_{A}}\delta\partial_{l}\phi _{A}\right]
=\delta\lbrack\delta L]=0.
\end{align}
The symplectic form is defined via the symplectic current by the same Eq.~(%
\ref{scalar}) as before. This form is closed because $J^{l}$ is closed, and
it is $\Sigma$-independent because $J^{l}$ is conserved\footnote{%
A subtlety arises if the theory in question is defined on a manifold with
spatial infinity. Strictly speaking, the conservation of $J^{l}$ only
implies that $\omega$ is invariant under variations which keep a part of $%
\Sigma$ near infinity fixed. A related problem is that the symplectic
current as defined by (\ref{_current}) can lead to a divergent symplectic
form. In this case a boundary term should be added to (\ref{scalar}) to
compensate the divergence. It turns out, however, that in all cases analysed
in this paper the symplectic form defined by the most natural expressions (%
\ref{scalar}), (\ref{_current}) is manifestly finite. We take it as an
indication that in our case no boundary terms are necessary, and that the
symplectic form is invariant also under shifts of $\Sigma$ at infinity.
However, since our plane wave computation in Sec.~\ref{PlaneWave} does not
match the expected answer, it is possible that the situation is not so
simple.}. It is slightly more complicated to show that in theories with
gauge symmetries the symplectic form so defined will be gauge invariant. For
specific equations it was shown in \cite{CW}, and a general argument can be
found in \cite{Zuckerman}, \cite{Wald}.

The sign and normalization of the definitions (\ref{scalar}), (\ref{_current}%
) can be checked by applying these formulas to the one-particle classical
mechanics, in which case they give, correctly
\begin{equation}
L=\int dt\left( \frac{1}{2}\dot{q}^{2}-V(q)\right) ,\qquad\omega
=J^{t}=\delta\dot{q}\wedge\delta q=\delta p\wedge\delta q.
\end{equation}

\subsection{Symplectic current of Type IIB SUGRA}

The LLM solutions satisfy equations of motion that can be derived from the
following action:
\begin{equation}
S=\frac{1}{2\kappa_{10}^{2}}\int d^{10}x\sqrt{-g}\left(
R-4|F_{5}|^{2}\right) ,   \label{IIB}
\end{equation}
We can proceed using this action if we impose the selfduality constraint $%
F_{5}=\ast F_{5}$ on the solutions. The presence of this constraint does not
modify the underlying symplectic form of the theory, which can be computed
from the action (\ref{IIB}){\footnote{%
In particular, one can perform the following analysis with the `actual' IIB
action \cite{actual} in which the selfduality constraint is imposed via an
auxiliary field.}}.

Using the general formulas of the previous subsection, the symplectic form
will be equal to%
\begin{equation}
\omega=\frac{1}{2\kappa_{10}^{2}}\int d\Sigma_{l}(J_{G}^{l}+J_{F}^{l}),
\label{sformIIB}
\end{equation}
where $J_{G}^{l}$ and $J_{F}^{l}$ are symplectic currents constructed from
the gravity and 5-form parts of the Langrangian using Eq.~(\ref{_current}).

To find the gravity current, we need first to remove the second derivatives
from the Einstein-Hilbert action, which is equivalent to adding the
Gibbons-Hawking boundary term \cite{GH}. The resulting first-derivative
action can be conveniently written in the so-called $\Gamma\Gamma-\Gamma%
\Gamma$ form (see e.g.\ \cite{LLII}):%
\begin{equation}
L=\sqrt{-g}g^{ik}[\Gamma_{il}^{m}\Gamma_{km}^{l}-\Gamma_{ik}^{l}\Gamma
_{lm}^{m}]\,.   \label{GammaGamma}
\end{equation}
(The sign is correct provided that the signature is mostly +.) We take the
inverse metric components $g^{mn}$ as our basic fields. Varying (\ref%
{GammaGamma}) with respect to $\partial_{l}g^{mn}$, it is easy to compute:
\begin{align}
\frac{\partial L}{\partial\partial_{l}g^{mn}} & =\sqrt{-g}%
[-\Gamma_{mn}^{l}+\delta_{\,(m}^{l}\Gamma_{n)k}^{k}+\frac{1}{2}%
g_{mn}(g^{ik}\Gamma_{ik}^{l}-g^{li}\Gamma_{ik}^{k})],  \notag \\
\frac{\partial L}{\partial\partial_{l}g^{mn}}\delta g^{mn} & =-\Gamma
_{mn}^{l}\delta\lbrack\sqrt{-g}g^{mn}]+\Gamma_{mn}^{n}\delta\lbrack\sqrt {-g}%
g^{lm}].
\end{align}
Thus we get
\begin{equation}
J_{G}^{l}=-\delta\Gamma_{mn}^{l}\wedge\delta\lbrack\sqrt{-g}%
g^{mn}]+\delta\Gamma_{mn}^{n}\wedge\delta\lbrack\sqrt{-g}g^{lm}],
\label{jgexp1}
\end{equation}
which is the old result of Crnkovi\'c and Witten \cite{CW}{\footnote{%
Actually, our result differs from that of \cite{CW} by an overall sign,
which can be traced to a different convention for the metric signature.}}
\footnote{%
It is stated in \cite{Wald} that this result is also contained, in a
disguised form, in \cite{Friedman}, but we were not able to locate it in
that paper.}. Notice that \cite{CW} defines the symplectic form as $\int
d\Sigma_{l}\sqrt{-g}J^{l},$ so that our symplectic currents differ from \cite%
{CW} by a factor of $\sqrt{-g}$.

To find the 5-form current$,$ we take the potentials $A_{|k_{1}\ldots k_{4}|}
$ $(F_{5}=dA)$ as our basic fields\footnote{%
The notation $|i_{1}\ldots i_{n}|$ means that the indices have to be
ordered: $i_{1}<\ldots<i_{n}.$ Thus we have $A=A_{|i_{1}\ldots
i_{4}|}dx^{i_{1}}\wedge\ldots\wedge dx^{i_{4}}=\frac{1}{4!}A_{i_{1}\ldots
i_{4}}dx^{i_{1}}\wedge\ldots\wedge dx^{i_{4}}.$ The same ordering is
assumed, e.g., in the summation in Eq.~(\ref{jfexp}).}. Applying (\ref%
{_current}), we get immediately
\begin{equation}
J_{F}^{l}=-8\,\delta(\sqrt{-g}F^{l|k_{1}\ldots k_{4}|})\,\delta
A_{|k_{1}\ldots k_{4}|}.   \label{jfexp}
\end{equation}
Now that we computed the symplectic current, we can simplify it using the
selfduality constraint, which can be written as
\begin{equation}
F^{l_{1}\ldots l_{5}}=\frac{1}{\sqrt{-g}}\varepsilon^{l_{1}\ldots
l_{5}|m_{1}\ldots m_{5}|}F_{|m_{1}\ldots m_{5}|},
\end{equation}
where $\varepsilon^{\ldots}$ is the flat 10-dimensional epsilon symbol. Thus
we have%
\begin{equation}
J_{F}^{l}=-8\,\varepsilon^{l|k_{1}\ldots k_{4}||m_{1}\ldots m_{5}|}\delta
F_{|m_{1}\ldots m_{5}|}\delta A_{|k_{1}\ldots k_{4}|}.   \label{jfexpsimp}
\end{equation}

\section{The $AdS$ background II -- CWZ approach}

\label{AdSTwo} We will\ now quantize small 1/2 BPS fluctuations about $%
AdS_{5}\times S^{5}$ using the CWZ method described in the previous section.
Thus we have to compute the symplectic currents (\ref{jgexp1}), (\ref{jfexp}%
) and integrate them to get the symplectic form (\ref{sformIIB}). The most
direct approach would be to use the LLM parametrization of the solutions in
terms of the functions $h,V_{i}$ etc. This approach is pursued in the next
section. In this section, we will use a hybrid approach. Namely, we will
once again use the modes of Kim \textit{et al.}\cite{Kim} which were related
to the LLM droplet deformations (\ref{droppar}) in Sec.~\ref{AdSOne}, Eqs.~(%
\ref{metpert}) and (\ref{deltaa}). However, we will not rely on the
effective action to perform the quantization, as we did in Sec. \ref{AdSOne}%
. Instead, we will use the Kim \textit{et al.} modes to evaluate the CWZ
symplectic currents and the symplectic form, which we then quantize.

In order to utilize the $t$-independence of the LLM solutions, we will make
the most natural choice of the hypersurface $\Sigma=\{t=const\}$ in (\ref%
{sformIIB}). Thus we only need to calculate $J^{t}$ to derive the symplectic
form $\omega$. The details of this calculation can be found in Appendix \ref%
{adsappendix}. The contribution from the gravity and 5-form current turn out
to be, respectively%
\begin{align}
\int_{t=\text{const}}J_{G}^{t} & =8\pi^{5}i\sum_{n\neq0}\frac{n^{2}-3|n|-8}{%
(|n|-1)(|n|+2)n}\,a_{n}\wedge a_{-n}\,,  \label{gravcurr} \\
\int_{t=\text{const}}J_{F}^{t} & =8\pi^{5}i\sum_{n\neq0}\frac{n^{2}+5|n|+4}{%
(|n|-1)(|n|+2)n}\,a_{n}\wedge a_{-n}\,.   \label{gaugecurr}
\end{align}
Notice that the coefficients $a_{n}$ of the Fourier series (\ref{droppar})
are treated here as one-forms on the LLM phase space. Substituting these
into (\ref{sformIIB}), we obtain the symplectic form%
\begin{equation}
\omega=\frac{8\pi^{5}i}{\kappa_{10}^{2}}\sum_{n\neq0}\frac{1}{n}%
\,a_{n}\wedge a_{-n}\,.   \label{adssympform}
\end{equation}
Inverting this symplectic form, we get the nonzero Poisson brackets:%
\begin{equation}
\{a_{m},a_{n}\}=-\frac{\kappa_{10}^{2}}{16\pi^{5}}in\delta_{m+n}\,.
\label{adscommrel}
\end{equation}
The commutators obtained from this bracket by the Dirac prescription $[\
,\,]=i\{\ ,\,\}$ agree with the result (\ref{acrs}) previously obtained by
the effective action method,\ and thus also with what we expect from the
fermion side, Eq.~(\ref{torepr}).

It should be noted that for $n=1$ Eqs.~(\ref{gravcurr}) and (\ref{gaugecurr}%
) are formally divergent. This divergence is due to the insufficiently fast
individual decay of $J_{G}^{t}$ and $J_{F}^{t}$ at large $\rho$ (see Eqs.~(%
\ref{firstterm}), (\ref{secondterm}), (\ref{Gaugecurr}) in Appendix \ref%
{adsappendix}). It is easy to check, however, that in the combined integral $%
\int J_{G}^{t}+J_{F}^{t}$ the divergence cancels, and the finite result of
this integration agrees with Eq.~(\ref{adssympform}), which is thus true
also for $n=1$.

\section{General droplets}

\label{CWforLLM}\label{PlaneWave}

In this section we will apply the CWZ method to write down the symplectic
form for the family of LLM configurations around an arbitrary droplet. In
this case, unlike for the $AdS_{5}\times S^{5}$ case we analyzed in Sec.~\ref%
{AdSTwo}, we don't have at our disposal a natural basis of linear
perturbations around this solution (analogous to the spherical harmonic
basis of Kim \textit{et al.}\cite{Kim} around $AdS_{5}\times S^{5}$). In
order to have general expressions, we will express the symplectic current in
terms of the LLM ansatz functions $h,V_{i}$ etc. As a concrete example, we
then apply these general expressions to the plane wave background.

Let us start from the gravitational piece of the symplectic current. Just
like in Sec.~\ref{AdSTwo}, we will use $\Sigma=\{t=const\}$ in (\ref%
{sformIIB}). Thus we only need to calculate $J^{t}$. Taking the general
expression (\ref{jgexp1}) and plugging in the expressions for the variations
of the metric and Christoffel symbols, we find the following expression:%
\begin{equation}
J_{G}^{t}=~y^{3}\left[ -\frac{1}{4}\delta(V_{i}\partial_{i}h^{-4})\
\delta(h^{4})+3\delta(V_{i}G_{,i})\delta G+\delta(h^{-4}W_{ij}V_{j})\ \delta
V_{i}-4\delta(\partial_{i}\ln h)\ \delta V_{i}\right]
\end{equation}
(where $i,j=1,2$, $W_{ij}\equiv\partial_{i}V_{j}-\partial_{j}V_{i}$). The
last term integrates to zero because $\partial_{i}V_{i}=0$, and can be
dropped. The first two terms add up nicely if one expresses $h$ and $G$ via $%
z$ from (\ref{h}), (\ref{G}). Remembering that $\left( \delta z\right) ^{2}=0
$, we get:
\begin{equation}
J_{G}^{t}=~y^{3}\left[ \frac{3/4+z^{2}}{\left( 1/4-z^{2}\right) ^{2}}%
\delta(V_{i}z_{,i})\delta z+\delta(h^{-4}W_{ij}V_{j})\ \delta V_{i}\right]%
\,.   \label{jgLLM}
\end{equation}

Similarly, using the expressions in (\ref{jfexpsimp}) and the form of the $%
F_{5}$ ansatz (\ref{ls})-(\ref{ll}), the $F_{5}$ current can be simplified
as follows:%
\begin{equation}
J_{F}^{t}=4\varepsilon^{abc}(\,\delta B_{a}\delta\tilde{F}_{bc}-\delta
\tilde{B}_{a}\delta F_{bc}\,).
\end{equation}
where $\varepsilon^{abc}$ is the flat space epsilon symbol in three
dimensions $y,x_{1},x_{2}.$ Since we will be working in the $B_{y}=\tilde{B}%
_{y}=0$ gauge, this expression can be simplified even further:
\begin{equation}
J_{F}^{t}=8\varepsilon^{ij}(\partial_{y}\delta B_{i}\,\delta\tilde{B}%
_{j}-\delta B_{i}\,\partial_{y}\delta\tilde{B}_{j}).   \label{jfLLM}
\end{equation}

Since the derived symplectic currents do not depend on the $S^{3},\tilde
{S}%
^{3}$ variables, the symplectic form (\ref{sformIIB}) can be simplified to
\begin{equation}
\omega=\frac{\left( 2\pi^{2}\right) ^{2}}{2\kappa_{10}^{2}}\int
dx_{1}dx_{2}dy\,(J_{G}^{t}+J_{F}^{t}).   \label{wLLM}
\end{equation}

Eqs.~(\ref{jgLLM}), (\ref{jfLLM}), (\ref{wLLM}) implicitly define the
symplectic form on the whole LLM class of solutions. We would like to make
an important remark here, namely that the above expressions in Eqs.~(\ref%
{jgLLM}), (\ref{jfLLM}) are the symplectic currents derived for the IIB
SUGRA action (\ref{IIB}) in the 10-dimensional bulk, with second derivatives
removed from the Einstein-Hilbert term. It is possible, as we noted in Sec.~%
\ref{CWmethod}, that this action has to be supplemented by extra boundary
terms, which would give an extra contribution to the symplectic currents.
For a discussion of this issue, in a slightly different context, see \cite%
{WaldZoupas}.

Explicit evaluation of the expression (\ref{wLLM}) in the general droplet
case is not easy, and is postponed to the future. Below we demonstrate how
to apply these expressions to evaluate the symplectic form around the plane
wave\footnote{%
Note that in this case, as with $AdS$, the spectrum of
fluctuations is known \cite{Metsaev} so in principle one could
also adopt an effective action approach as in section
\ref{Effective}. Since our interest is only to demonstrate
applicability of the CWZ formalism to a second example, we don't
pursue this here.}.

The plane wave background corresponds to a droplet which covers the entire
lower half-plane, i.e. $Z(x_{1},x_{2})=\frac{1}{2}sign\,x_{2}$. The various
functions specifying the solution are given by:%
\begin{align}
z & =\frac{x_{2}}{2r},\qquad r\equiv\sqrt{x_{2}^{2}+y^{2}} ,  \notag \\
V_{1} & =-\frac{1}{2r},\quad V_{2}=0\,,  \label{pw} \\
h^{-2} & =2r\,.  \notag
\end{align}

Now we look at small perturbations around this configuration, where the
boundary of the droplet is deformed from $x_{2}=0$ into $x_{2}=\varepsilon
(x_{1})$ \footnote{%
As we are looking at small perturbations, this description is enough, and we
do not consider situations where the droplet boundary winds such that it has
a few $x_{2}$ values for a specific $x_{1}$, or where the topology of the
boundary is changed and some disconnected droplets appear.}. Note that this
deformation is area preserving if $\int_{-\infty}^{\infty
}dx_{1}\varepsilon(x_{1})=0$. Making such a deformation takes us into a new
solution with new functions $z(x_{1},x_{2},y)\,,V_{i}(x_{1},x_{2},y)$. The
corresponding variations can be found from (\ref{z}), (\ref{V}); to first
order in $\varepsilon(x_{1})$ they are given by:%
\begin{align}
\delta z(x_{1},x_{2},y) & =-\frac{y^{2}}{\pi}\int dx_{1}^{\prime}\frac{%
\varepsilon(x_{1}^{\prime})}{[(x_{1}-x_{1}^{\prime})^{2}+r^{2}]^{2}}%
\equiv-y^{2}I_{1} ,  \notag \\
\delta V_{1}(x_{1},x_{2},y) & =-\frac{x_{2}}{\pi}\int dx_{1}^{\prime}\frac{%
\varepsilon(x_{1}^{\prime})}{[(x_{1}-x_{1}^{\prime})^{2}+r^{2}]^{2}}%
=-x_{2}I_{1},  \label{dzdv} \\
\delta V_{2}(x_{1},x_{2},y) & =\frac{1}{\pi}\int dx_{1}^{\prime}\frac {%
(x_{1}-x_{1}^{\prime})\varepsilon(x_{1}^{\prime})}{[(x_{1}-x_{1}^{\prime
})^{2}+r^{2}]^{2}}\equiv I_{2} .  \notag
\end{align}
We see that these expressions have the form of convolutions in the $x_{1}$
variable. It thus makes sense to perform a Fourier transform w.r.t. $x_{1}$
(keeping the other two variables intact), which will turn these convolutions
into products. Thus for a general function $f(x_{1},x_{2},y)$ we will have:%
\begin{equation*}
f(x_{1},x_{2},y)=\int\frac{dp}{2\pi}\,e^{-ipx_{1}}\tilde{f}(p,x_{2},y),\qquad%
\tilde{f}(p,x_{2},y)\equiv\int dx_1\,e^{ipx_{1}}f(x_{1},x_{2},y).\qquad
\end{equation*}

The $\delta h$ can be expressed via $\delta z$ from (\ref{h}). All these
variations should be substituted into (\ref{jgLLM}), and eventually into the
integral (\ref{wLLM}). The dependence on $p$ can be removed from the
integrand by rescaling $r|p|\rightarrow r$, with the remaining $x_{2},y$
integral giving rise to a constant prefactor. A very similar computation is
carried out for the 5-form part of the current. These computations are
detailed in Appendix \ref{Appendix}. The final result is that the symplectic
form is of the form:
\begin{equation}
\omega\propto\int\frac{dp}{2\pi}\frac{i}{p}\,\tilde{\varepsilon} (p)\wedge%
\tilde{\varepsilon}(-p).  \label{final}
\end{equation}From this symplectic form we get the Poisson brackets
$\{\tilde{\varepsilon }(p),\tilde{\varepsilon}(p^{\prime})\}$ and,
by the Dirac prescription, the commutators. Here too we find that
the commutator of $\varepsilon(x_{1})$ and
$\varepsilon(x_{1}^{\prime})$ constitute a $U(1)$ Kac-Moody
algebra, and
are thus of the same form as one expects based on the fermion analysis (\ref%
{ffcomm}) (to see this, one has to apply (\ref{ffcomm}) around a circular
droplet of very large radius $R$. In this limit, we can identify the plane
wave variable $x_{1}$ with the arc length measured along the boundary of the
droplet, and $\varepsilon(x_{1})$ \ with the perturbation of the droplet
radius). The commutators which we find (see Appendix \ref{Appendix}) seem to
be a factor 2 of what one expects from (\ref{ffcomm}). We believe that this
numerical mismatch is due to extra boundary terms which should be added to (%
\ref{wLLM}), as we remarked before. This issue is currently under
investigation, and we hope to report on it in a future publication.

\section{Conclusions and discussion}

\label{Conclusions}

In this paper we have considered the LLM family of solutions and set up a
general framework for its quantization. In particular, using the CWZ method,
we have performed the quantization in the case of the $AdS_5\times S^5$
background and demonstrated its consistency with a more direct effective
action approach. This has permitted us to construct a quantum Hilbert space
associated to the LLM solutions that precisely matches that of the
corresponding fermion picture at large $N$. This result allows us to
identify the spacetime $x_{1},x_{2}$ plane with the Fermi liquid phase
space, in agreement with the original LLM proposal. Moreover, we have
derived general expressions for the symplectic form that can, in principle,
be applied to more complicated droplets.

The outlook for future developments is as follows. First of all, one should
explore the issue of boundary terms to be added to the general symplectic
form (\ref{wLLM}) derived in Sec.~\ref{CWforLLM}. As discussed in Sec.~\ref%
{PlaneWave}, this should also be used to complete our quantization around
the plane wave background\footnote{%
One could also try to perform a higher-order expansion around the
half-plane droplet. Such a computation could be useful to
determine if a boundary term in the symplectic form is indeed
missing. Note that the lowest-order plane wave computation of
Sec.~\ref{PlaneWave} is not suitable for such a check, since every
lowest-order term in (\ref{jgLLM}), (\ref{jfLLM}) integrates
separately to an expression of the same functional form.}. It will
also be of interest to generalize our techniques so that they may
be easily applied to droplets with more complicated shapes%
\footnote{In this respect, see \cite{Dhar} for a discussion of how
to quantize fermion droplets correspoding to multiple-valued
$r(\phi)$.}
as well as nontrivial topologies%
\footnote{%
For instance, in the case of an annulus surrounding the origin we expect to
get two identical $U(1)$ Kac-Moody algebras, one for each boundary, as the
only coupling between the degrees of freedom on each boundary is through the
total area conservation rule.}.

Another interesting course of work is the application of these methods to
additional families of supergravity solutions that arise in various
contexts. One example is the family of \cite{Diana} with nonzero axion and
dilaton, in which the solutions correspond to changing the fermions
participating in the Quantum Hall Effect into fractional statistics
particles. It would be interesting to reproduce this behavior from the
gravity point of view. Another example is the family of 1/2 BPS M2 brane
solutions of 11-dimensional SUGRA described by LLM (see also subsequent work
\cite{bak}).

Another important class of solutions to which one might apply these methods
are the 1/4 BPS D1-D5 solutions with angular momentum constructed in \cite%
{LMM}. As these are supersymmetric, we once again expect a quantization, at
least to the order considered in our case, to have meaning even within the
context of the full theory. The results would be very interesting \cite{our}
and could have implications in the context of D1-D5 black hole entropy as
well as suggestions made by Mathur \textit{et al.} regarding the relation of
these solutions to the D1-D5 black hole \cite{Mathur}.

Finally, it would also be interesting to consider going beyond the
semiclassical approximation in supergravity and working at finite $N$. Of
course, as soon as $N$ is taken away from infinity we have to deal with
corrections from string theory. As a result, proceeding along this direction
seems very difficult. It is possible, though, that the correspondence to
fermions contains hints as to how one might proceed since the fermion
picture essentially tells us what the correct quantization should look like.
In fact, it seems to indicate that whatever $1/N$ corrections arise must
somehow conspire to yield a free theory in the appropriate description. It
would be fascinating to understand this in detail.

\section*{Note added in proof}

We have recently found what caused the factor 2 mismatch in
section 6. The reason is that the four-form potentials are
singular if the axial gauge (\ref{gauge}) is used. The singularity
occurs at the $y = 0$ plane, where the spheres $S^3$ and $\tilde
S^3$, whose volume forms enter the potentials, shrink to zero. To
compensate for this singularity, a boundary term located at $y =
0$ should be added to the symplectic form. More details can be
found in \cite{MR}, where the general droplet case is also
analyzed in full.

\acknowledgments
We would like to thank O.~Aharony, J.~de Boer, A.~Dhar, J.~Esole, N.~Iizuka,
O.~Lunin, J.~Maldacena, L.~Motl, A.~Naqvi, I.~Papadimitriou, S.~Raju,
K.~Schoutens, A.~Shomer, K.~Skenderis, M.~Smedback, and especially
S.~Minwalla for useful discussion. We would also like to thank A.~Dhar and
S.~Minwalla for collaboration at an early stage of this work. L.G., K.P.,
and J.M. would like to thank the Tata Institute of Fundamental Research for
providing a stimulating research atmosphere during the beginning stages of
this work. L.M. would like to thank the Weizmann Institute of Science for
hospitality during the final period of this work. The work of L.G., J.M.,
and K.P. is supported by DOE grant DE-FG01-91ER40654. The work of J.M.\ is
also supported in part by an NSF Graduate Research Fellowship. The work of
L.M.\ and V.R.\ is supported by Stichting FOM.

\appendix

\section{Evaluation of the CWZ currents around $AdS_{5}\times S^{5}$}

\label{adsappendix} In this appendix, we provide details for the
computations reported in Sec.~\ref{AdSTwo}. Thus we will compute the
symplectic currents (\ref{jgexp1}), (\ref{jfexp}) and the symplectic form (%
\ref{sformIIB}) for fluctuations about the $AdS_{5}\times S^{5}$ background
parameterized by (\ref{snyn}), (\ref{metpert}), (\ref{deltaa}). We proceed
by brute force, computing the nonvanishing components of the various tensors
appearing in (\ref{jgexp1}), (\ref{jfexp}).

Let $\alpha,\beta,\gamma$ ($a,b,c)$ denote the three angles of the $S^{3}(%
\tilde{S}^{3})$ in (\ref{glob}). The volume forms of the $S^{3}$'s will be
written in terms of these variables as%
\begin{equation}
d\Omega_{3}=\sin^{2}\alpha\sin\beta\,d\alpha\wedge d\beta\wedge d\gamma
\,\,,\qquad\qquad d\tilde{\Omega}_{3}=\sin^{2}a\sin b\,da\wedge db\wedge
dc\,.   \label{s3volforms}
\end{equation}

The only nonzero components of $\delta\left[ g^{mn}\sqrt{-g}\right] $ are
given by
\begin{align}
\delta\left[ g^{tt}\sqrt{-g}\right] & =-\sum_{n\neq0}2a_{n}(|n|+1)e^{in(%
\tilde{\phi}+t)}\frac{\tanh^{2}\rho}{\cosh^{2}\rho}\sqrt{-\bar {g}}\,%
\mathcal{C}_{n}\,,  \notag \\
\delta\left[ g^{t\rho}\sqrt{-g}\right] & =-\sum_{n\neq0}\frac{2ina_{n}}{|n|}%
(|n|+1)e^{in(\tilde{\phi}+t)}\frac{\tanh\rho}{\cosh^{2}\rho}\sqrt {-\bar{g}}%
\,\mathcal{C}_{n}\,,  \notag \\
\delta\left[ g^{\rho\rho}\sqrt{-g}\right] & =\sum_{n\neq0}2a_{n}(|n|+1)e^{in(%
\tilde{\phi}+t)}\frac{\sqrt{-\bar{g}}\,\mathcal{C}_{n}}{\cosh ^{2}\rho}\,, \\
\delta\left[ g^{\alpha\alpha}\sqrt{-g}\right] &
=\sum_{n\neq0}2a_{n}(|n|+1)e^{in(\tilde{\phi}+t)}\frac{\sqrt{-\bar{g}}\,%
\mathcal{C}_{n}}{\sinh^{2}\rho}\,,  \notag \\
\delta\left[ g^{\beta\beta}\sqrt{-g}\right] &
=\sum_{n\neq0}2a_{n}(|n|+1)e^{in(\tilde{\phi}+t)}\frac{\sqrt{-\bar{g}}\,%
\mathcal{C}_{n}}{\sinh^{2}\rho\sin^{2}\alpha}\,,  \notag \\
\delta\left[ g^{\gamma\gamma}\sqrt{-g}\right] &
=\sum_{n\neq0}2a_{n}(|n|+1)e^{in(\tilde{\phi}+t)}\frac{\sqrt{-\bar{g}}\,%
\mathcal{C}_{n}}{\sinh^{2}\rho\sin^{2}\alpha\sin^{2}\beta}\,,  \notag
\end{align}
where%
\begin{equation*}
\sqrt{-\bar{g}}=\sinh^{3}\rho\cosh\rho\sin^{3}\theta\cos\theta\sin^{2}\alpha%
\sin\beta\sin^{2}a\sin b
\end{equation*}
is the square root of the determinant of the unperturbed $AdS_{5}\times S^{5}
$ metric and%
\begin{equation}
{\mathcal{C}}_{n}\equiv\left( \frac{\cos\theta}{\cosh\rho}\right) ^{|n|}\,.
\label{cndef}
\end{equation}
Using these, we find that the only nonzero $\delta\Gamma_{mn}^{t}$ which
contribute to the first term of $J_{G}^{t}$ are:
\begin{align}
\delta\Gamma_{tt}^{t} & =-\frac{1}{2}\sum_{n\neq0}\frac{ina_{n}}{|n|}e^{in(%
\tilde{\phi}+t)}\left( |n|+1\right) \left( |n|+4-2\frac {|n|+2}{\cosh^{2}\rho%
}\right) \,{\mathcal{C}}_{n}\,,  \notag \\
\delta\Gamma_{t\rho}^{t} & =\frac{1}{2}\sum_{n\neq0}a_{n}e^{in(\tilde{\phi }%
+t)}\left( |n|+1\right) \left( |n|-2\frac{|n|+2}{\cosh^{2}\rho}\right)
\tanh\rho\,\,{\mathcal{C}}_{n}\,,  \notag \\
\delta\Gamma_{\rho\rho}^{t} & =-\frac{1}{2}\sum_{n\neq0}\frac{ina_{n}}{|n|}%
e^{in(\tilde{\phi}+t)}\,\left( |n|+1\right) \left( 3|n|-2\frac {|n|+2}{%
\cosh^{2}\rho}\right) \frac{\,{\mathcal{C}}_{n}}{\cosh^{2}\rho}\,, \\
\delta\Gamma_{\alpha\alpha}^{t} & =-\frac{1}{2}\sum_{n\neq0}\frac{ina_{n}}{%
|n|}e^{in(\tilde{\phi}+t)}\left( |n|+1\right) (|n|-4)\,\tanh^{2}\rho\,\,{%
\mathcal{C}}_{n}\,,  \notag \\
\delta\Gamma_{\beta\beta}^{t} & =-\frac{1}{2}\sum_{n\neq0}\frac{ina_{n}}{|n|}%
e^{in(\tilde{\phi}+t)}\left( |n|+1\right) (|n|-4)\,\tanh^{2}\rho
\sin^{2}\alpha\,\,{\mathcal{C}}_{n}\,,  \notag \\
\delta\Gamma_{\gamma\gamma}^{t} & =-\frac{1}{2}\sum_{n\neq0}\frac{ina_{n}}{%
|n|}e^{in(\tilde{\phi}+t)}\left( |n|+1\right) (|n|-4)\,\tanh^{2}\rho
\sin^{2}\alpha\sin^{2}\beta\,\,{\mathcal{C}}_{n}\,,  \notag
\end{align}
while the only nonzero $\delta\Gamma_{mp}^{m}$ that contribute to the second
term of $J_{G}^{t}$ are:
\begin{align}
\delta\Gamma_{mt}^{m} & =\sum_{n\neq0}ina_{n}e^{in(\tilde{\phi}+t)}(|n|+1)\,{%
\mathcal{C}}_{n}\,, \\
\delta\Gamma_{m\rho}^{m} & =-\sum_{n\neq0}a_{n}e^{in(\tilde{\phi}%
+t)}|n|(|n|+1)\tanh\rho\,{\mathcal{C}}_{n}\,.
\end{align}
It is now a trivial matter to compute $J_{G}^{t}$. We find it easier to
present $\int\,d\tilde{\phi}\,J_{G}^{t}$ rather than $J_{G}^{t}$ itself.
From the first term, we get%
\begin{multline}
\int\,d\tilde{\phi}\,\delta\Gamma_{mp}^{t}\wedge\delta\left[ \sqrt{-g}g^{mp}%
\right] \\
=\sum_{n\neq0}\frac{4\pi in(|n|+1)^{2}{\mathcal{C}}_{n}^{\,2}}{%
|n|\cosh^{6}\rho}\left[ 7+|n|-3|n|\cosh(2\rho)+\cosh(4\rho)\right] \sqrt{-%
\bar{g}}\left( a_{n}\wedge a_{-n}\right)   \label{firstterm}
\end{multline}
Performing the remaining integrals, we find%
\begin{equation}
\int_{t=\text{const}}\,\delta\Gamma_{mp}^{t}\wedge\delta\left[ \sqrt {-g}%
g^{mp}\right] =-\sum_{n\neq0}\frac{2\pi i(2\pi^{2})^{2}\left(
n^{2}-|n|-8\right) }{(|n|-1)(|n|+2)n}a_{n}\wedge a_{-n}   \label{firallints}
\end{equation}
For the second term contributing to $J_{G}^{t}$, we have%
\begin{equation}
\int\,d\tilde{\phi}\,\delta\Gamma_{mp}^{p}\wedge\delta\left[ \sqrt{-g}g^{mt}%
\right] =-\sum_{n\neq0}\frac{8\pi in\,{\mathcal{C}}_{n}^{2}\tanh ^{2}\rho}{%
\cosh^{2}\rho}\left( 1+|n|\right) ^{2}\sqrt{-\bar{g}}\left( a_{n}\wedge
a_{-n}\right)   \label{secondterm}
\end{equation}
Performing the remaining integrals here, we find%
\begin{equation}
\int_{t=\text{const}}\,\delta\Gamma_{mp}^{p}\wedge\delta\left[ \sqrt {-g}%
g^{mt}\right] =-\sum_{n\neq0}\frac{4\pi i(2\pi^{2})^{2}|n|}{(|n|-1)(|n|+2)n}%
\left( a_{n}\wedge a_{-n}\right)   \label{secallints}
\end{equation}
Combining (\ref{firallints}) and (\ref{secallints}), we obtain precisely
Eq.~(\ref{gravcurr}) of Sec.~\ref{AdSTwo}.

We now turn to the 5-form current. From the expression (\ref{deltaa}) we can
compute the perturbation of the 5-form itself%
\begin{multline}
\delta F_{5} =\sum_{n\neq0}a_{n}{\mathcal{C}}_{n}e^{in(\tilde{\phi}%
+t)}\left( \frac{|n|+1}{2|n|}\right) \times \\
\times \biggl[\biggl(|n|(|n|-4)\cosh\rho \sinh^{3}\rho\,dt\wedge
d\rho-n^{2}\sinh^{2}\rho\,\tanh\rho\,d\rho\wedge d\tilde{\phi}%
+n^{2}\sinh^{4}\rho\tan\theta\,dt\wedge d\theta \\
-in|n|\sinh^{4}\rho\,dt\wedge d\tilde{\phi}-in|n|\sinh^{2}\rho\tanh\rho\tan%
\theta \,d\rho\wedge d\theta\biggr)\wedge d\Omega_{3} \\
+\biggl(|n|(|n|+4)\cos\theta\sin^{3}\theta\,d\theta\wedge d\tilde{\phi }%
-n^{2}\sin^{2}\theta\,\tan\theta\,dt\wedge d\theta-n^{2}\sin^{4}\theta
\tanh\rho\,d\rho\wedge d\tilde{\phi} \\
\quad+in|n|\sin^{4}\theta\,dt\wedge d\tilde{\phi}-in|n|\sin^{2}\theta
\tan\theta\tanh\rho\,d\rho\wedge d\theta\biggr)\wedge d\tilde{\Omega}_{3}%
\biggr]   \label{deltaf}
\end{multline}
The variations that contribute to (\ref{jfexp}) (in this computation we
haven't made use of the simplified form (\ref{jfexpsimp})) are as follows%
\begin{equation}
\begin{split}
\delta A_{\rho\alpha\beta\gamma} & =-\sum_{n\neq0}ina_{n}{\mathcal{C}}%
_{n}e^{in(\tilde{\phi}+t)}\left( \frac{|n|+1}{2|n|}\right)
\sinh^{2}\rho\tanh\rho\sin^{2}\alpha\sin\beta \\
\delta A_{\tilde{\phi}abc} & =\sum_{n\neq0}|n|a_{n}{\mathcal{C}}_{n}e^{in(%
\tilde{\phi}+t)}\left( \frac{|n|+1}{2|n|}\right) \sin^{4}\theta\sin^{2}a\sin
b \\
\delta A_{\theta abc} & =\sum_{n\neq0}ina_{n}{\mathcal{C}}_{n}e^{in(\tilde{%
\phi}+t)}\left( \frac{|n|+1}{2|n|}\right) \sin^{2}\theta
\tan\theta\sin^{2}a\sin b
\end{split}%
\end{equation}%
\begin{equation}
\begin{split}
\delta\left[ \sqrt{-g}F^{t\rho\alpha\beta\gamma}\right] & =-\sum_{n\neq 0}%
\frac{a_{n}{\mathcal{C}}_{n}}{2}e^{in(\tilde{\phi}+t)}(|n|+1)(|n|+4)\cos
\theta\sin^{3}\theta\sin^{2}a\sin b \\
\delta\left[ \sqrt{-g}F^{t\tilde{\phi}abc}\right] & =-\sum_{n\neq0}\frac{%
a_{n}{\mathcal{C}}_{n}}{2}e^{in(\tilde{\phi}+t)}in(|n|+1)\tanh\rho
\sinh^{2}\rho\sin^{2}\alpha\sin\beta\tan\theta \\
\delta\left[ \sqrt{-g}F^{t\theta abc}\right] & =\sum_{n\neq0}\frac {a_{n}{%
\mathcal{C}}_{n}}{2}e^{in(\tilde{\phi}+t)}|n|(|n|+1)\tanh\rho\sinh
^{2}\rho\sin^{2}\alpha\sin\beta
\end{split}%
\end{equation}
With this, we find that%
\begin{equation}
\int\,d\tilde{\phi}\,\delta A_{|\ldots|}\wedge\delta\left[ \sqrt {-g}%
F^{t|\ldots|}\right] =\sum_{n\neq0}\frac{i\pi n(1+|n|)^{2}{\mathcal{C}}%
_{n}^{\,2}[1+|n|+\cos(2\theta)]}{|n|\cos^{2}\theta\cosh^{2}\rho}\sqrt {-\bar{%
g}}\left( a_{n}\wedge a_{-n}\right)   \label{Gaugecurr}
\end{equation}
Performing the remaining integrals, we reproduce Eq.~(\ref{gaugecurr}) of
Sec.~\ref{AdSTwo}.

\section{Evaluation of the CWZ currents around the plane wave}

\label{Appendix} In this appendix we detail the derivation of the symplectic
form around the plane wave background, outlined in Sec.~\ref{PlaneWave}. Let
us start from the gravitational part of the symplectic form. We need
expressions for all variations entering (\ref{jgLLM}) in terms of $I_{1,2}.$
Such expressions can be obtained using (\ref{h})-(\ref{G}) and (\ref{dzdv}).
We have:%
\begin{align}
\delta h^{-4} & =-16r^{3}x_{2}I_{1}\,,  \notag \\
\delta(V_{i}z_{,i}) & =\frac{y^{2}}{2r^{3}}\left( I_{2}+r^{2}\partial
_{1}I_{1}\right) \,,  \label{dh} \\
\delta W_{12} & =\partial_{1}I_{2}+\partial_{2}(x_{2}I_{1})\,.  \notag
\end{align}

We use the Fourier transforms of the integrals $I_{1},I_{2}$ given by:
\begin{align}
\tilde{I}_{1}(p,x_{2},y) & =\frac{1+r|p|}{2r^{3}}e^{-r|p|}\tilde {\varepsilon%
}(p)  \notag \\
\tilde{I}_{2}(p,x_{2},y) & =\frac{ip}{2r}e^{-r|p|}\tilde{\varepsilon}(p)
\label{I12}
\end{align}
Then the first term in (\ref{jgLLM}) gives the following contribution to the
integral in (\ref{wLLM}):%
\begin{align}
& \int dx_{1}dx_{2}dy\,\,y^{3}\left[ \frac{4r^{2}(3r^{2}+x_{2}^{2})}{y^{4}}%
\right] \frac{y^{2}}{2r^{3}}\left( I_{2}+r^{2}\partial_{1}I_{1}\right)
\left( -y^{2}I_{1}\right)  \notag \\
& =\int\frac{dp}{2\pi}\int dx_{2}dy\,\,(3r^{2}+x_{2}^{2})\frac{-2y^{3}}{r}%
\left[ \tilde{I}_{2}(p)+r^{2}(-ip)\tilde{I}_{1}(p)\right] \tilde{I}_{1}(-p)
\notag \\
& =\int\frac{dp}{2\pi}\int dx_{2}dy\,\left( 3+\frac{x_{2}^{2}}{r^{2}}\right)
\frac{y^{3}}{2r^{2}}\,(1+r|p|)e^{-2r|p|}ip|p|\,\tilde{\varepsilon }(p)\tilde{%
\varepsilon}(-p)  \notag \\
& =\int dx_{2}dy\,\left( 3+\frac{x_{2}^{2}}{r^{2}}\right) \frac{y^{3}}{2r^{2}%
}\,(1+r)e^{-2r}\times\Omega\equiv c_{1}\Omega\,,  \notag \\
& \Omega \equiv\int\frac{dp}{2\pi}\frac{i}{p}\tilde{\varepsilon}(p)\,\tilde{%
\varepsilon}(-p),   \label{before}
\end{align}
where we rescaled variables ($x_{2},y)\rightarrow(x_{2}$, $y)/|p|$ to remove
all dependence on $p$ from the $x_{2},y$ integral, which became a constant
prefactor. This prefactor can be easily evaluated:%
\begin{equation}
c_{1}=\int_{0}^{\pi}d\phi\int_{0}^{\infty}dr\,r\,(3+\cos^{2}\phi)\frac {%
r\sin^{3}\phi}{2}\,(1+r)e^{-2r}=\frac{4}{3}.   \label{before1}
\end{equation}
(the angular integral is from $0$ to $\pi$ because of $y>0$).

Using that $V_{2}=0$ for the plane wave, the second term in (\ref{jgLLM})
can be simplified as follows:%
\begin{multline}
y^{3}\left[ \delta(h^{-4}W_{12}V_{2})\delta
V_{1}-\delta(h^{-4}W_{12}V_{1})\delta V_{2}\right] \\
=-y^{3}\left[ 2W_{12}h^{-4}\delta V_{1}\delta V_{2}+W_{12}V_{1}\delta
h^{-4}\,\delta V_{2}+h^{-4}V_{1}\delta W_{12}\,\delta V_{2}\right] .
\end{multline}
The first two terms here cancel because of $2h^{-4}\delta V_{1}+V_{1}\delta
h^{-4}=0$, as one can check using (\ref{pw}), (\ref{dzdv}), (\ref{dh}). The
remaining term gives the following contribution to the integral in (\ref%
{wLLM}):%
\begin{align}
& \int dx_{1}dx_{2}dy\,\,2ry^{3}\left[ \partial_{1}I_{2}+%
\partial_{2}(x_{2}I_{1})\right] I_{2}  \notag \\
& =\int\frac{dp}{2\pi}\int dx_{2}dy\,2ry^{3}\left[ -ip\tilde{I}%
_{2}(p)+\partial_{2}(x_{2}\tilde{I}_{1}(p))\right] \tilde{I}_{2}(-p)  \notag
\\
& =\int\frac{dp}{2\pi}\int dx_{2}dy\,\,\frac{-y^{3}}{2r^{3}}\left[
1+r|p|+r^{2}p^{2}-\frac{x_{2}^{2}}{r^{2}}\left( 3+3r|p|+r^{2}p^{2}\right) %
\right] e^{-2r|p|}ip\,\tilde{\varepsilon}(p)\tilde{\varepsilon}(-p)  \notag
\\
& \equiv c_{2}\Omega\,,  \notag \\
& c_{2} =-\int dx_{2}dy\,\frac{y^{3}}{2r^{3}}\,\left[ 1+r+r^{2}-\frac {%
x_{2}^{2}}{r^{2}}\left( 3+3r+r^{2}\right) \right] e^{-2r}=-\frac{1}{3}\,,
\label{after1}
\end{align}
where we have used the same method of computing the integral as in (\ref%
{before}), (\ref{before1}).

Let us now proceed with the gauge field part of the symplectic current, (\ref%
{jfLLM}). Expressions for the gauge fields were given in (\ref{gauge}); we
have to find their variations. Using (\ref{dzdv}) it is easy to show that
\begin{align}
\delta \frac{y^{2}V_{1}}{{\frac{1}{2}}\pm z}& =2r^{2}x_{2}I_{1}\mp
2r^{3}I_{1}\,, \\
\delta \frac{y^{2}V_{2}}{{\frac{1}{2}}\pm z}& =2r^{2}I_{2}\mp 2rx_{2}I_{2}\,.
\end{align}%
We also have%
\begin{align}
\delta \frac{1}{\pi }\frac{x_{2}}{x^{2}+y^{2}}\ast Z& =-\frac{x_{2}}{\pi }%
\int dx_{1}^{\prime }\frac{\varepsilon (x_{1}^{\prime })}{%
(x_{1}-x_{1}^{\prime })^{2}+r^{2}}\equiv -x_{2}I_{3\,,} \\
\delta \frac{1}{\pi }\frac{x_{1}}{x^{2}+y^{2}}\ast Z& =-\frac{1}{\pi }\int
dx_{1}^{\prime }\frac{(x_{1}-x_{1}^{\prime })\varepsilon (x_{1}^{\prime })}{%
(x_{1}-x_{1}^{\prime })^{2}+r^{2}}\equiv -I_{4}\,,
\end{align}%
with the corresponding Fourier transforms in $x_{1}$ given by%
\begin{align}
\tilde{I}_{3}(p,x_{2},y)& =\frac{1}{r}e^{-r|p|}\tilde{\varepsilon}(p)\,,
\notag \\
\tilde{I}_{4}(p,x_{2},y)& =i\,sign\,p\,e^{-r|p|}\tilde{\varepsilon}(p)\,.
\label{I34}
\end{align}%
We thus have:%
\begin{align}
-4\delta B_{1}& =\left( 2r^{2}x_{2}I_{1}-x_{2}I_{3}\right)
+2r^{3}I_{1}\equiv a_{1}+b_{1}\,,  \notag \\
-4\delta \tilde{B}_{1}& =\left( 2r^{2}x_{2}I_{1}-x_{2}I_{3}\right)
-2r^{3}I_{1}=a_{1}-b_{1}\,,  \notag \\
-4\delta B_{2}& =(2r^{2}I_{2}+I_{4})+2rx_{2}I_{2}\equiv a_{2}+b_{2}\,, \\
-4\delta \tilde{B}_{2}& =(2r^{2}I_{2}+I_{4})-2rx_{2}I_{2}=a_{2}-b_{2}\,.
\notag
\end{align}%
The purpose of introducing this notation is that in (\ref{jfLLM}) many
cross-terms cancel:
\begin{equation}
J_{F}^{t}=\left( \frac{\partial a_{2}}{\partial y}b_{1}-a_{2}\frac{\partial
b_{1}}{\partial y}\right) -\left( \frac{\partial a_{1}}{\partial y}%
b_{2}-a_{1}\frac{\partial b_{2}}{\partial y}\right) .  \label{purp}
\end{equation}%
The Fourier transforms of $a_{i},b_{i}$ are easily found using (\ref{I12}), (%
\ref{I34}); we have:%
\begin{align}
\tilde{a}_{1}(p)& =x_{2}|p|e^{-r|p|}\,,  \notag \\
\tilde{b}_{1}(p)& =(1+|p|r)e^{-r|p|}\,,  \notag \\
\tilde{a}_{2}(p)& =i\,sign\,p(1+|p|r)e^{-r|p|}=i\,sign\,p\,\,\tilde{b}%
_{1}(p)\,, \\
\tilde{b}_{2}(p)& =ix_{2}pe^{-r|p|}=i\,sign\,p\,\,\tilde{a}_{1}(p)\,.  \notag
\end{align}%
Because of these proportionalities, substituting these expressions in $\int
J_{F}^{t}$ results in a total cancellation (multiplication by $i\,sign\,p$
will commute with the $y$ derivatives, and the integrands corresponding to
both terms in the right-hand side of (\ref{purp}) will vanish on the Fourier
side).

Thus we conclude that the gauge part does not contribute to the symplectic
form around the plane wave (this is unlikely to be true in the general
case). The final answer is given by adding the gravitational part
contributions (\ref{before1}), (\ref{after1}); we have%
\begin{equation}
\omega=\frac{(2\pi^{2})^{2}}{2\kappa_{10}^{2}}\Omega= \frac{2\pi^{4}}{%
\kappa_{10}^{2}}\int\frac{dp}{2\pi}\frac{i}{p}\,\tilde{\varepsilon}(p)\tilde{%
\varepsilon}(-p).
\end{equation}
This symplectic form implies the commutators:
\begin{equation}
[\tilde{\varepsilon}(p),\tilde{\varepsilon}(p^{\prime})]=-\frac {%
\kappa_{10}^{2}}{2\pi^{3}}p\,\delta(p+p^{\prime}).
\end{equation}
From this it is easy to get the commutators in the coordinate
representation:
\begin{equation}
\lbrack\varepsilon(x_{1}),\varepsilon(x_{1}^{\prime})]=-i\frac{%
\kappa_{10}^{2}}{4\pi^{4}}\delta^{\prime}(x_{1}-x_{1}^{\prime}).
\label{ecomm}
\end{equation}

We would like to compare the last equation with the finite droplet result (%
\ref{ffcomm}). For this we have to apply (\ref{ffcomm}) around a circular
droplet of very large radius $R$. In this limit, we can identify the plane
wave variable $x_{1}$ with the arc length measured along the boundary of the
droplet:%
\begin{equation}
x_{1}=-R\phi
\end{equation}
(the sign is chosen to preserve orientation) and $\varepsilon(x_{1})$ with
the perturbation in the droplet radius. Thus we have%
\begin{equation}
f^{2}(\phi)\approx R^{2}+2R\varepsilon(x_{1}).
\end{equation}
Using $\delta^{\prime}(\phi-\phi^{\prime})=-R^{2}\delta^{%
\prime}(x_{1}-x_{1}^{\prime})$, we see that (\ref{ffcomm}) in the limit $%
R\rightarrow \infty$ gives the result which is of precisely the same
functional form as (\ref{ecomm}), but contains an extra factor 1/2 in the
right-hand side. We suspect that the reason for this mismatch, as we
mentioned in Sec.~\ref{PlaneWave}, is that there should be a boundary term
added to the symplectic form, which would give an equal contribution as the
bulk term we have just evaluated.


\begin{thebibliography}{99}
\bibitem{AdSCFT} J.~M.~Maldacena, ``The large N limit of superconformal
field theories and supergravity,"Adv.\ Theor.\ Math.\ Phys.\
\textbf{2}, 231 (1998) [Int.\ J.\ Theor.\ Phys.\ \textbf{38}, 1113
(1999)]
[arXiv:hep-th/9711200]. 
\newline
S.~S.~Gubser, I.~R.~Klebanov and A.~M.~Polyakov, ``Gauge theory
correlators from non-critical string theory," Phys.\ Lett.\ B
\textbf{428}, 105 (1998)
[arXiv:hep-th/9802109]. 
\newline
E.~Witten, ``Anti-de Sitter space and holography," \ Adv.\ Theor.\
Math.\ Phys.\ \textbf{2}, 253 (1998) [arXiv:hep-th/9802150].

\bibitem{Corley}  S.~Corley, A.~Jevicki and S.~Ramgoolam,  ``Exact
correlators of giant gravitons from dual N = 4 SYM theory,''
Adv.\ Theor.\ Math.\ Phys.\ \textbf{5}, 809 (2002)
[arXiv:hep-th/0111222].

\bibitem{Berenstein} D.~Berenstein, ``A toy model for the AdS/CFT
correspondence,'' JHEP \textbf{0407}, 018 (2004)
[arXiv:hep-th/0403110].

\bibitem{LLM} H.~Lin, O.~Lunin and J.~Maldacena, ``Bubbling AdS space and
1/2 BPS geometries,'' JHEP \textbf{0410}, 025 (2004)
[arXiv:hep-th/0409174].

\bibitem{Mandal} G.~Mandal, ``Fermions from half-BPS supergravity,''
arXiv:hep-th/0502104. 

\bibitem{Kim} H.~J.~Kim, L.~J.~Romans and P.~van Nieuwenhuizen, ``The Mass
Spectrum Of Chiral $\mathcal{N}=2$ $D = 10$ Supergravity On
$S^{5}$,''
Phys.\ Rev.\ D \textbf{32}, 389 (1985). 

\bibitem{Shiraz} S.~M.~Lee, S.~Minwalla, M.~Rangamani and N.~Seiberg,
``Three-point functions of chiral operators in $D = 4$,
$\mathcal{N} = 4$ SYM at large $N$,'' Adv.\ Theor.\ Math.\ Phys.\
\textbf{2}, 697 (1998)
[arXiv:hep-th/9806074]. 

\bibitem{ADM} R.~Arnowitt, S.~Deser and C.~W.~Misner, ``The Dynamics Of
General Relativity,'' in \emph{Gravitation: an introduction to
current research}, Ed.\ L.~Witten (Wiley 1962), p.227
[arXiv:gr-qc/0405109].

\bibitem{CW} \v{C}.~Crnkovi\'c and E.~Witten, ``Covariant Description Of
Canonical Formalism In Geometrical Theories,'' in \emph{Three
hundred years of gravitation}, Eds.\ S.W.~Hawking and W.~Israel
(Cambridge University Press, 1987), p.676.

\bibitem{Zuckerman} G.~J.~Zuckerman, ``Action Principles And Global
Geometry," in \emph{Mathematical Aspects Of String Theory}, San
Diego 1986, Proceedings, Ed.\ S.-T.~Yau (Worls Scientific, 1987),
p.259.

\bibitem{Misner} C.~W.~Misner, ``Minisuperspace,'' in \emph{Magic Without
Magic: John Archibald Wheeler}, Ed.\ J.~R.~Klauder (Freeman, San
Francisco 1972), p. 441.

\bibitem{Wheeler} J.~A.~Wheeler, ``Geometrodynamics and the issue of the final state," in \emph{Relativity, Groups and Topology},
Eds.\ C.~DeWitt and B.~DeWitt (Gordon and Breach, New York, 1964)

\bibitem{DeWitt} B.~S.~DeWitt, ``Quantum Theory Of Gravity. I. The Canonical
Theory,'' Phys.\ Rev.\ \textbf{160}, 1113 (1967).

\bibitem{Kuchar} K.~Kucha\v{r}, ``Canonical Quantization Of Cylindrical
Gravitational Waves,'' Phys.\ Rev.\ D \textbf{4}, 955 (1971).
\newline
G.~A.~Mena Marug\'an and M.~Montejo, ``Quantization of pure
gravitational plane waves,'' Phys.\ Rev.\ D \textbf{58}, 104017
(1998)
[arXiv:gr-qc/9806105]. 

\bibitem{KMM} J.~Kinney, J.~Maldacena and S.~Minwalla, to appear.

\bibitem{KlebanovRev} I.~R.~Klebanov, ``TASI lectures: Introduction to the
AdS/CFT correspondence,'' arXiv:hep-th/0009139.

\bibitem{Sunny} N.~Itzhaki and J.~McGreevy, ``The large $N$ harmonic
oscillator as a string theory,'' Phys.\ Rev.\ D \textbf{71},
025003 (2005)
[arXiv:hep-th/0408180]. 

\bibitem{Ber2} D.~Berenstein, ``A matrix model for a quantum Hall droplet
with manifest particle-hole symmetry,'', Phys.\ Rev.\ D
\textbf{71} (2005)
085001 [arXiv:hep-th/0409115]. 

\bibitem{Wen} X.-G.~Wen, ``Topological orders and Edge excitations in FQH
states", arXiv:cond-mat/9506066.

\bibitem{Dhar}
  A.~Dhar,
  ``Bosonization of non-relativstic fermions in 2-dimensions and collective
  field theory,''
  arXiv:hep-th/0505084.

\bibitem{Poly} A.~P.~Polychronakos, ``Chiral actions from phase space
(quantum Hall) droplets,'' Nucl.\ Phys.\ B \textbf{705}, 457
(2005)
[arXiv:hep-th/0408194]. 

\bibitem{Haldane} F.~D.~M.~Haldane, ``Stability of Chiral Luttinger Liquids
and Abelian Quantum Hall States,'' Phys.\ Rev.\ Lett.\ 74,
\textbf{2090} (1995) [arXiv:cond-mat/9501007].

\bibitem{actual} G.~Dall'Agata, K.~Lechner and D.~P.~Sorokin, ``Covariant
actions for the bosonic sector of D = 10 IIB supergravity,''
Class.\ Quant.\ Grav.\ \textbf{14} (1997) L195
[arXiv:hep-th/9707044].

\bibitem{Witten} E.~Witten, ``Interacting Field Theory Of Open
Superstrings,'' Nucl.\ Phys.\ B \textbf{276}, 291 (1986).

\bibitem{WZW} M.~f.~Chu, P.~Goddard, I.~Halliday, D.~I.~Olive and
A.~Schwimmer, ``Quantization of the Wess-Zumino-Witten model on a
circle,''
Phys.\ Lett.\ B \textbf{266}, 71 (1991). 
\newline
Y.~Nutku, ``Lagrangian approach to integrable systems yields new
symplectic structure for KdV,'' in \textit{Integrable hierarchies
and modern physical theories}, Chicago 2000, Proceedings, Eds.\
H.~Aratyn and A.~Sorin (Kluwer,
2001), p.203 [arXiv:hep-th/0011052]. 
\newline
J.~Lucietti, ``Canonical quantization of a massive particle on
AdS(3),'' JHEP \textbf{0305} (2003) 017 [arXiv:hep-th/0303228].

\bibitem{Wald} J.~Lee and R.~M.~Wald, ``Local Symmetries And Constraints,''
J.\ Math.\ Phys.\ \textbf{31}, 725 (1990). 
\newline
B.~Julia and S.~Silva, ``On covariant phase space methods,''
arXiv:hep-th/0205072. 

\bibitem{GH} G.~W.~Gibbons and S.~W.~Hawking,  ``Action Integrals And
Partition Functions In Quantum Gravity,''  Phys.\ Rev.\ D
\textbf{15}, 2752
(1977).  

\bibitem{LLII} L.~D.~Landau and E.~M.~Lifshitz, \emph{The Classical Theory
of Fields}, (Butterworth-Heinemann, 1980).

\bibitem{Friedman} J.~L.~Friedman, ``Generic Instability of Rotating
Relativistic Stars," Comm.\ Math.\ Phys.\ \textbf{62}, 247 (1978).

\bibitem{WaldZoupas} R.~M.~Wald and A.~Zoupas, ``A General Definition of
``Conserved Quantities" in General Relativity and Other Theories
of Gravity,'' Phys.\ Rev.\ D \textbf{61}, 084027 (2000)
[arXiv:gr-qc/9911095].

\bibitem{Metsaev}  R.~R.~Metsaev and A.~A.~Tseytlin,  ``Exactly solvable
model of superstring in plane wave Ramond-Ramond  background,''
Phys.\ Rev.\ D \textbf{65}, 126004 (2002)  [arXiv:hep-th/0202109].

\bibitem{Diana} J.~T.~Liu, D.~Vaman and W.~Y.~Wen, ``Bubbling 1/4 BPS
solutions in type IIB and supergravity reductions on $S^{n}\times
S^{n}$,''
arXiv:hep-th/0412043. 
\newline
J.~T.~Liu and D.~Vaman, ``Bubbling 1/2 BPS solutions of minimal
six-dimensional supergravity,'' arXiv:hep-th/0412242.

\bibitem{bak} D.~Bak, S.~Siwach and H.~U.~Yee, ``1/2 BPS geometries of M2
giant gravitons,'' arXiv:hep-th/0504098. 

\bibitem{LMM} O.~Lunin, J.~Maldacena and L.~Maoz, ``Gravity solutions for
the D1-D5 system with angular momentum,'' arXiv:hep-th/0212210.

\bibitem{our} L.~Maoz, V.~S.~Rychkov and A.~Shomer, in progress.

\bibitem{Mathur} S.~D.~Mathur ``The fuzzball proposal for black holes: An
elementary review,'' arXiv:hep-th/0502050. 
\newline
O.~Lunin and S.~D.~Mathur, ``AdS/CFT duality and the black hole
information paradox,'' Nucl.\ Phys.\ B \textbf{623}, 342 (2002)
[arXiv:hep-th/0109154].

\bibitem{MR}
  L.~Maoz and V.~S.~Rychkov,
  ``Geometry quantization from supergravity: The case of 'Bubbling AdS',''
  arXiv:hep-th/0508059.


\end{thebibliography}
\end{document}